\documentclass[a4paper,11pt]{article}
\usepackage{geometry}
\usepackage[colorlinks=true,urlcolor=black,linkcolor=blue,citecolor=blue,bookmarks=false,breaklinks]{hyperref}
\usepackage{amsfonts,amsmath,amssymb}
\usepackage{subcaption}
\usepackage{graphicx}
\usepackage[font=small,labelfont=bf]{caption}
\usepackage{cite}
\usepackage[page,toc,titletoc,title]{appendix}
\usepackage{url}
\usepackage[OT1]{fontenc}
\usepackage{authblk}
\usepackage{fancyhdr}
\usepackage{booktabs}

\graphicspath{{figures/}}

\renewcommand*{\thepage}{\footnotesize\arabic{page}}
\providecommand{\keywords}[1]{\small{\textbf{Keywords:} #1}}

\newcommand{\avg}[1]{\langle #1 \rangle}
\newcommand{\corr}[2]{C_{#1}^{(\text{#2})}}

\title{\bf Spatial Correlation Analysis of Traffic Flow on Parallel Motorways in Germany}

\author{Sebastian Gartzke\thanks{sebastian.gartzke@uni-due.de}, Shanshan Wang, Thomas Guhr and Michael Schreckenberg}
\affil{\textit{Faculty of Physics, University Duisburg--Essen, Lotharstra\ss e 1, 47048 Duisburg, Germany}}

\date{\today}
 
\begin{document}
\maketitle

\begin{abstract}
\noindent
With the widely used method of correlation matrix analysis, this study reveals the change of traffic states on parallel motorways in North Rhine-Westphalia, Germany. In terms of
the time series of traffic flow and velocity, we carry out a quantitative analysis in correlations and reveal a high level of strongly positive traffic flow correlation and rich structural features in the corresponding correlation matrices. The strong correlation is mainly ascribed to the daily time evolution of traffic flow during the periods of rush hours and non-rush hours. In terms of free flow and congestion, the structural features are able to capture the average traffic situation we derive from our data. Furthermore, the structural features in correlation matrices for individual time periods corroborate our results from the correlation matrices regarding a whole day. The average correlations in traffic flows and velocities over all pairwise sections disclose the traffic behavior during each individual time period. Our contribution uncovers the potential application of correlation analysis on the study of traffic networks as a complex system.

\vspace{\baselineskip} \noindent \keywords{correlation matrix, time series analysis, vehicular traffic, traffic state,  complex system}
\end{abstract}


\noindent\rule{\textwidth}{1pt}
\vspace*{-1cm}
{\setlength{\parskip}{0pt plus 1pt} \tableofcontents}
\noindent\rule{\textwidth}{1pt}

\frenchspacing
\clearpage
\section{Introduction}
\label{sec1}
Vehicular traffic is a keystone for mobility and economic infrastructure in today's society. The worldwide increasing number of vehicles \cite{UmBund} forces traffic networks of large urban agglomerations to the limits of capacity, resulting in congestion and increased travel times. Keeping these perturbations to a minimum is an important and desirable task for maintaining the functionality of these networks. Therefore, studying the dynamics of these many-body systems is a difficult but crucial task for effective traffic management and control. 

The complex \cite{Ladyman2013} nature of vehicular traffic is well reflected by the non-stationary characteristics of the traffic variables flow, velocity and density \cite{Treiber2010}. The behavior of traffic flow and the emergence of different traffic states, i.e. free flow and congestion \cite{Kerner_2009}, has been studied empirically \cite{Neubert_1999, Tilch_2000, Lee_2000, Kerner_2002_E, Schoernhof_2004, Bertini_2005} as well as with a large variety of modeling approaches and simulations \cite{NaSch_1992, Schadschneider_1993, Schreckenberg_1995, Barlovic_1998, Kerner_2002, Knospe_2002, Schadschneider_2010}. With this study we follow a new approach for the investigation of vehicular traffic networks and non-stationary traffic flow time series. 

Analyzing correlation matrices of time series is a common method for studying the temporal evolution of complex systems~\cite{Muennix_2012}. In the research field of financial markets, in which stocks can be considered as interacting components, this method has particularly proven itself. By considering a financial market as a whole system several studies based on empirical data uncovered the possibility to associate structural correlations with market states, through which the system passes in time\cite{Muennix_2012,Chetalova_2015,Rinn_2015,Stepanov_2015,Heckens_2020}. Identifying similar structures by clustering correlation matrices allows to identify market states at different time periods of varying length. Therefore it is possible to study the temporal evolution of the market and identify possible state transitions. This is a powerful tool for investigating and understanding the non-stationary characteristics of these systems. 

A recent study \cite{Wang2020} successfully transfers the concepts explained to studying empirical traffic data of stationary loop detectors on German motorways. By clustering correlation matrices of traffic flow position series, it is possible to identify quasi-stationary states. Considering each time evolution, some of these states can be associated with workdays, other with holidays (meaning weekends and public holidays). Diagonal block structures of the workday states in time periods of rush hours and non-rush hours indicate correlations of congested and free flow states of traffic. 

The present study aims at uncovering further potential for studying road networks with methods of correlation analysis. Here we focus on longer ranged spatial correlations. To this end, empirical traffic data collected by stationary loop detectors on two German motorways, running parallel, are investigated. The existence of structural features in the correlation matrices of non-stationary traffic flow and velocity time series is shown, analyzed quantitatively and connected with the traffic situation on the studied motorways. Furthermore the correlation matrices and the average correlation of individual time periods, e.g. rush hours and non-rush hours, are investigated. 

This paper is organized as follows. Section \ref{sec2} introduces the studied data set. Section \ref{sec3} describes the calculation of correlation matrices and introduces conventions and traffic flow fundamentals. The analysis of the correlation matrices is conducted in section \ref{sec4}. Our results are concluded in section \ref{sec5}.

\section{Data}
\label{sec2}
\begin{figure}[tbp]
\centering
	\begin{subfigure}[b]{1\textwidth}
		\caption{}
		\includegraphics[width=1\textwidth]{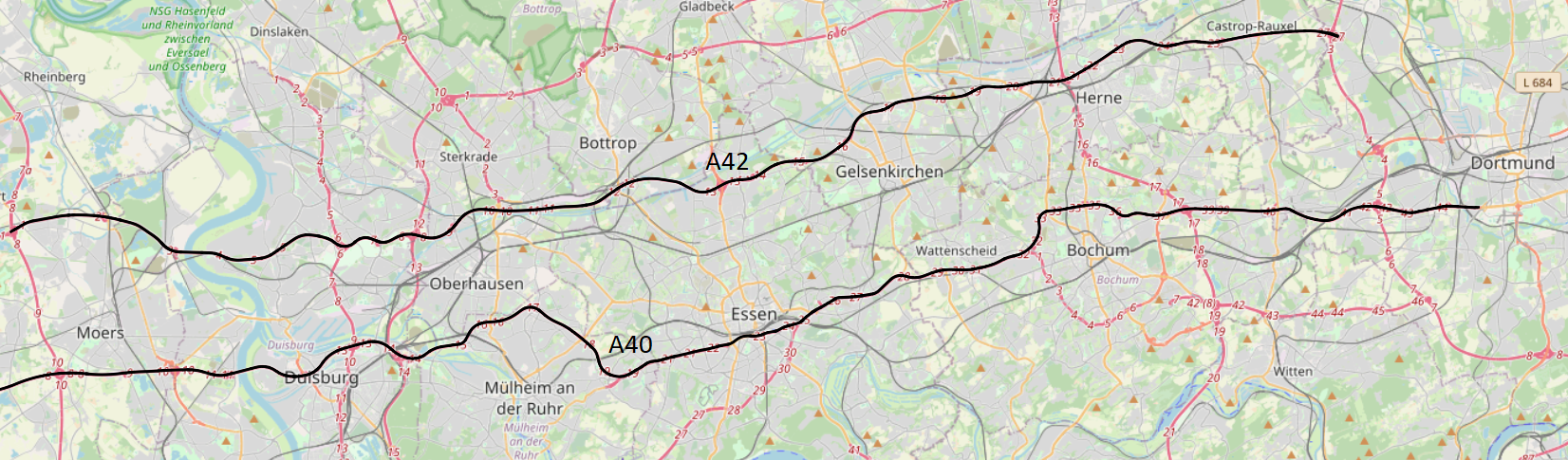}
	\end{subfigure}

	\begin{subfigure}[b]{1\textwidth}
		\caption{}
		\includegraphics[width=1\textwidth]{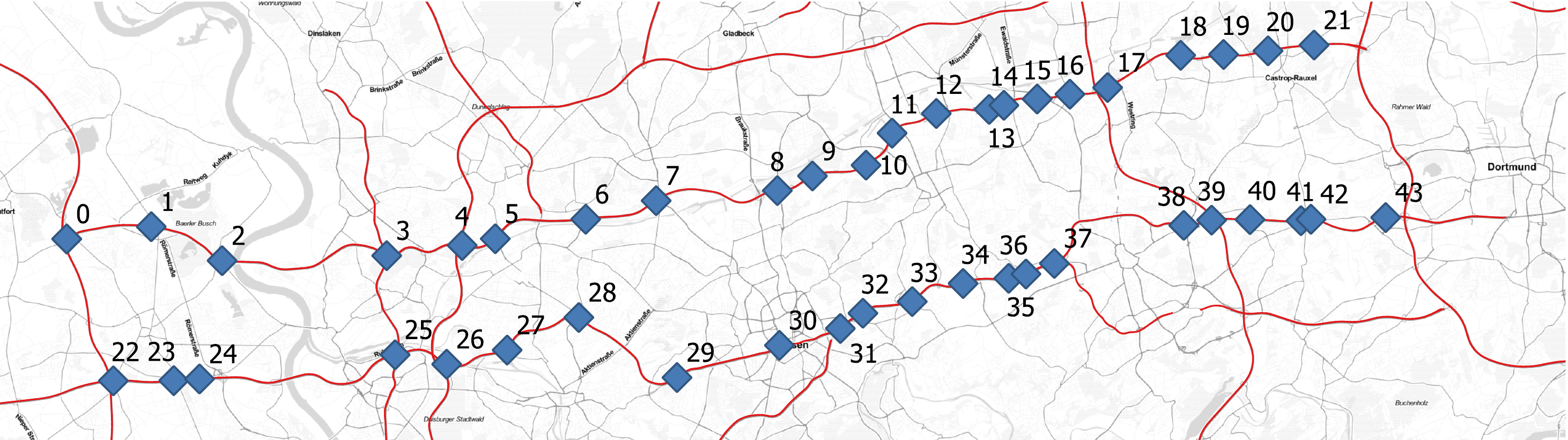}
	\end{subfigure}
		
		\caption{(a) Routes of the motorways A40 and A42 (black lines) through North Rhine-Westpalia (Germany). (b) Locations of traffic detectors (blue markers) along motorway A40 (cross-sections~0-21) and A42 (cross-sections~22-43). At each cross-section the traffic situation on all lanes in north-east direction is observed. The map data for both maps and the map tiles in (a) are provided by OpenStreetMap \textcopyright OpenStreetMap Contributors and are licensed under ODbL v1.0 \cite{OSML,OpDbLi}. The map tiles in (b) are made by Stamen Design licensed under CC BY 3.0 \cite{StaDes}.}
		\label{fig_1_map}
\end{figure}

The Ruhrgebiet (Ruhr area) in North Rhine-Westphalia (NRW), Germany, is one of the largest urban agglomerations in Europe \cite{Eurostat}. Traffic congestion is a major problem, especially on motorways used by commuters and trucks regularly. The most prominent example is the motorway A40. Passing through the whole Ruhr area, A40 connects major cities like Duisburg, Mülheim a.d. Ruhr, Essen, Bochum, and Dortmund strung like pearls on a cord. Figure \ref{fig_1_map}(a) shows the route of A40 through the Ruhr area. A40 is one of the most trafficked and congested motorways in Germany \cite{Adac_stau}. The motorway segment between the city of Essen and Dortmund exhibits more traffic jams than any other part of the German motorway network.

The traffic data used in this study were obtained from inductive loop detectors along the motorways A40 and A42 in north-east (NE) direction in 2017. For each of these motorways the data of 22 road cross-sections is used, covering both motorways on an equal footing. The 44 cross-sections along both motorways observing traffic in NE direction are located as shown in Fig. \ref{fig_1_map}(b). At each cross-section the detectors gather information on the traffic situation on each lane, such as traffic flow rate $q$ and average speed $v$, over time intervals of one minute. The aggregated data obtained are in units of vehicles/h and km/h, respectively, and distinguish between cars and trucks. For the sake of completeness Appendix~\ref{ap_data_ex} contains an example of the investigated time series data. To enhance the quality of the data set, all days on which any of the detectors did not produce traffic flow data are sorted out. In addition, public holidays are excluded, leaving a total of $N=218$ days in 2017 for the study. This includes $N_{\text{wd}}=158$ workdays and $N_{\text{we}}=60$ weekend days.

Similar to reference \cite{Wang2020} the data of all available lanes of each section are combined into one effective lane. For processing our data we consider the hydrodynamic relation
\begin{equation}
	\rho = \frac{q}{v},
	\label{eq_hydro}	
\end{equation}
which is usually applied to determine densities $\rho$ from empirical traffic data. Notice that Eq. \eqref{eq_hydro} has to be considered as an approximation for the calculation of $\rho$ with data gathered by stationary loop detectors because the determination of spatial quantities from temporal quantities can be biased with systematic errors \cite{Treiber2010}. Regarding the definition of density, Eq.\eqref{eq_hydro} only yields exact results for the spatial mean speed and not for the time mean speed measured at stationary loop detectors. In case of the time mean speed the true density is generally underestimated by Eq. \eqref{eq_hydro}. For the purpose of our work we consider the approximation with Eq. \eqref{eq_hydro} sufficient. Therefore, for each cross-section~$k$ and each lane~$l$ a traffic density
\begin{equation}
	\rho_{kl}^{(m)}(t) =  \frac{q_{kl}^{(m)}(t)}{v_{kl}^{(m)}(t)}, 
	\label{eq1}
\end{equation}
with $m = \text{car},\text{tr}$, is calculated for cars and trucks separately based on the measured traffic flows~$q_{kl}^{(m)}(t)$ and velocities~$v_{kl}^{(m)}(t)$. The total traffic flow~$q_{kl}(t)$ and density~$\rho_{kl}(t)$ of cars and trucks for each lane are given by the sums
\begin{align}
	q_{kl}(t) &= q_{kl}^{(\text{car})}(t) + q_{kl}^{(\text{tr})}(t), \label{eq2} \\ 	
	\rho_{kl}(t) &= \rho_{kl}^{(\text{car})}(t) + \rho_{kl}^{(\text{tr})}(t). 	\label{eq3}
\end{align}
The effective traffic flow~$q_k(t)$ and density~$\rho_k(t)$ for each cross-section corresponds to the sum over all lanes and the velocities~$v_k(t)$ result from
\begin{equation}
	v_k(t) = \frac{q_{k}(t)}{\rho_{k}(t)} = \frac{\sum_{l} q_{kl}(t)}{\sum_{l} \rho_{kl}(t)}.
	\label{eq4}
\end{equation}
After calculating the effective traffic flows and velocities the time series are aggregated to a longer time interval by summing over all traffic flows and averaging over all velocities for each cross-section. Based on the results in reference \cite{Wang2020} we choose an aggregation time of 15 minutes.

\section{Methods}
\label{sec3}

\subsection{Correlation matrices}
\label{sec31}
The daily time series of traffic flow and velocity of $K=44$ cross-sections along the motorways A40 and A42 are analyzed. For each day we calculate the $K \times T$ data matrices
\begin{equation}
G =	\begin{bmatrix}
		G_1(1) & \cdots & G_1(T) \\
		\vdots &		& \vdots \\
		G_k(1) & \cdots & G_k(T) \\
		\vdots &		& \vdots \\
		G_K(1) & \cdots & G_K(T)
	\end{bmatrix},
\label{eq_data_matrix}
\end{equation}
where $G$ either contains the traffic flow or the velocity data. The $k$-th row of $G$ is the time series~$G_k(t)$ of cross-section~$k$ where $k=1,\dots,K$. All~$G_k(t)$ have a length of $T=96$, where each~$t=1,\dots,T$ corresponds to a time interval of 15 minutes from 00:00 to 23:59. All time series $G_k(t)$ are standardized to mean zero and unit standard deviation, resulting in a normalized data matrix $M$ with entries
\begin{equation}
	M_k(t) = \frac{G_k(t) - \avg{G_k(t)}_T}{\sigma_k}, \quad k=1, \dots, K, \ t=1,\dots,T.
	\label{eq:standardize}
\end{equation}
The time average is written as
\begin{equation}
	\avg{G_k(t)}_T = \frac{1}{T} \sum_{t=1}^{T} G_k(t)
	\label{eq:average}
\end{equation}
and the standard deviation is given by
\begin{equation}
	\sigma_k = \sqrt{\avg{G^{2}_k(t)}_T - \avg{G_k(t)}^{2}_T}.
	\label{eq:deviation}
\end{equation}
The $K \times K$ Pearson correlation coefficients
\begin{equation}
	C_{kl} = \frac{1}{T} \sum_{t=1}^{T} M_k(t) M_l(t)
	\label{eq_corr_coef}
\end{equation}
are the elements of the correlation matrix
\begin{equation}
	C = \frac{1}{T} M M^{\dagger}
	\label{eq_corr_mat}
\end{equation}
capturing all $K$ detectors. The superscript $\dagger$ indicates the matrix transpose. The rows of the data matrices and therefore the ones of the correlation matrices are arranged according to Fig.~\ref{fig_1_map}(b), hence the first half of rows correspond to the cross-sections on A42 and the second half to the ones on A40. Because of this arrangement each resulting correlation matrix can be considered as consisting of four smaller square matrices: The upper left square matrix contains the correlations of the sections on A42, the lower right square matrix contains the ones of the sections on A40. The off diagonal matrices are identical (due to symmetry) and contain the correlations between the sections of the two motorways.

We consider time series of traffic flow and velocity. Thus, matrices containing traffic flow data are denoted with a subscript~$q$ while matrices containing velocity data are denoted with a subscript~$v$. Furthermore, traffic flow correlation matrices averaged over all workdays (weekends) are denoted by~$C^{\text{(wd)}}_q$ ($C^{\text{(we)}}_q$). Averaged velocity correlation matrices are denoted analogously.

\subsection{Traffic flow fundamentals}
\label{sec32}
\begin{figure}[htbp]
	\begin{center}
		\includegraphics[width=0.9\textwidth]{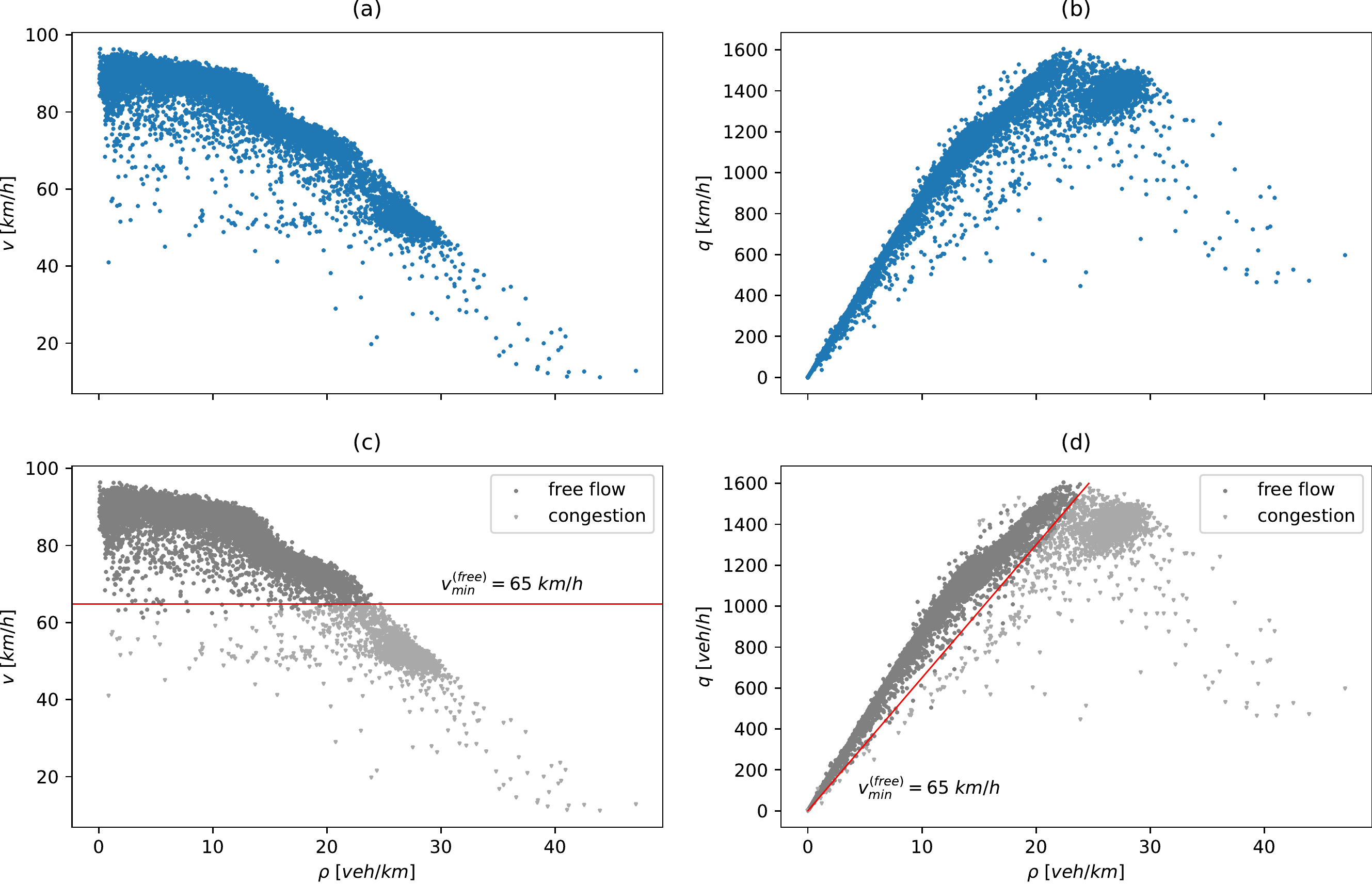}
		\caption{Example of an empirical velocity-density (a) and flow-density (b) diagram. For the separation of free flow and congested traffic phases k-menas clustering is applied on the velocity-density plane (c). The corresponding minimum free flow velocity $v_{\text{min}}^{\text{free}}$ is indicated by a red line. The resulting clusters from k-mean algorithm can be mapped onto the flow-density plane (d) where the slope of the red line correspond to $v_{\text{min}}^{\text{free}}$. The diagrams contain the data points of 218 days measured at cross-section 32.}
		\label{fig_2_funda_ex}
	\end{center}
\end{figure}

\noindent In the simplest case traffic flow theory distinguishes between two states (or phases) of vehicular traffic, namely free traffic flow (free flow) and congested traffic \cite{Kerner_2009,Schadschneider_2010}. A free flow traffic phase is usually characterized by a small vehicle density~$\rho$ and a low interaction rate of the individual traffic participants. For small enough densities the vehicle interaction rate is low. Several studies, e.g. \cite{Li_2017,Krbalek_2016}, found that headways between vehicles follow Poissonian distribution for very small densities. Under normal circumstances the vehicle interactions only influence the headways between succeeding vehicles which does not force the drivers to change their velocity significantly. Therefore the traffic participants may travel with their desired maximum velocity (which can be limited by certain conditions e.g. speed-limitations). With increasing traffic density the flow rate~$q$ (flow) also increases in free flow phases. This increase of flow, however, is limited because the interaction rate of vehicles also increases with further growing density. As a consequence the average velocity~$v$ of the traffic participants decreases due to vehicle interactions which are therefore no longer negligible. 

The limit of free flow is given at a maximum flow~rate~$q_{\text{max}}^{\text{(free)}}$ and density~$\rho_{\text{max}}^{\text{(free)}}$, respectively. Considering the hydrodynamic relation in Eq.~\eqref{eq_hydro} the average velocity in free flow phases has a minimum value~\cite{Kerner_2009} of 
\begin{equation}
	v_{\text{min}}^{\text{(free)}} = \frac{q_{\text{max}}^{\text{(free)}}}{\rho_{\text{max}}^{\text{(free)}}}.
	\label{eq_vmin_free}
\end{equation}
For a high enough density increase during a free flow phase the average velocity may drop abruptly due to a higher vehicle interaction rate. Such a decrease in average velocity during free flow is considered as the occurrence of congestion. A traffic phase in which the average velocity is smaller than the minimum velocity for free flow is defined as congested traffic \cite{Kerner_2009}. 

The relation between flow, density, and velocity in different traffic phases can be visualized by considering the fundamental relation (in literature often referred to as fundamental diagram) of empirical traffic data. Figure.~\ref{fig_2_funda_ex} shows two different representations of the fundamental relation, namely the velocity-density diagram (Fig.~\ref{fig_2_funda_ex}(a)) and the flow-density diagram (Fig.~\ref{fig_2_funda_ex}(b)). For the purpose of our work we identify free flow and congested traffic phases within our data for each cross-section and determine the corresponding minimum free flow velocities $v_{\text{min}}^{\text{(free)}}$. For the identification of the traffic phases we use a clustering approach with the k-means algorithm \cite{Wang2020,Kianfar_2013,Kianfar_2010}. K-means clustering is a centroid-based clustering method where the number of centroids $k$ is preset and the initial centroids are selected randomly. For each cross-section we apply k-means clustering on the velocity-density plane \cite{Kianfar_2013} containing the data points of all $N=218$ days. According to the number of traffic states we aim to identify, we preset the number of centroids to $k=2$. Figure~\ref{fig_2_funda_ex}(c) shows an example for the identification of free flow and congestion resulting from k-means clustering. In Figure~\ref{fig_2_funda_ex}(d) we map the resulting clusters onto the flow-density plane. For the determination of the minimum free flow velocities $v_{\text{min}}^{\text{(free)}}$ we identify the maximum velocity $v_{\text{max}}^{\text{(cong,cl)}}$ in the congestion cluster for each cross-section. 

\section{Spatial correlations}
\label{sec4}

\subsection{Traffic situation on A40 and A42}
\label{sec41}

\begin{figure}[htbp]
	\begin{center}
		\includegraphics[width=0.6\textwidth]{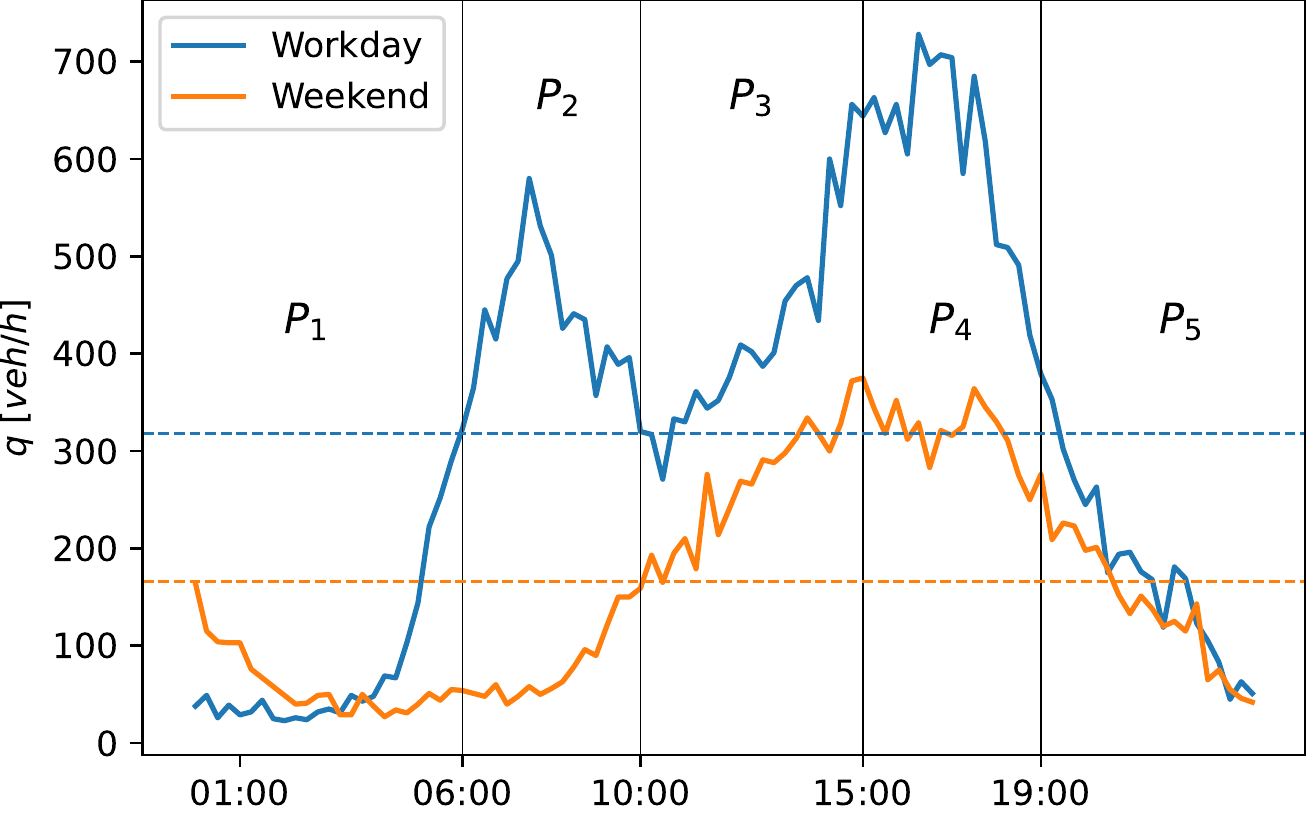}
		\caption{Examples for the time evolution of traffic flow on workdays and weekends. Vertical black lines separate a workday into five time periods: The night time ($P_1$), the morning and afternoon rush hours ($P_2$ and $P_4$), the midday ($P_3$) and the evening ($P_5$). Colored horizontal dashed lines mark the mean values of each time series. The data shown for workdays and weekends was measured at cross-section~5 on Tuesday, February 7th, 2017 and on Sunday, February 12, 2017, respectively, and is aggregated as described in section \ref{sec2}.}
		\label{fig_3_tagesgang}
	\end{center}
\end{figure}

\begin{figure}[tbp]
	\begin{center}
		\includegraphics[width=0.8\textwidth]{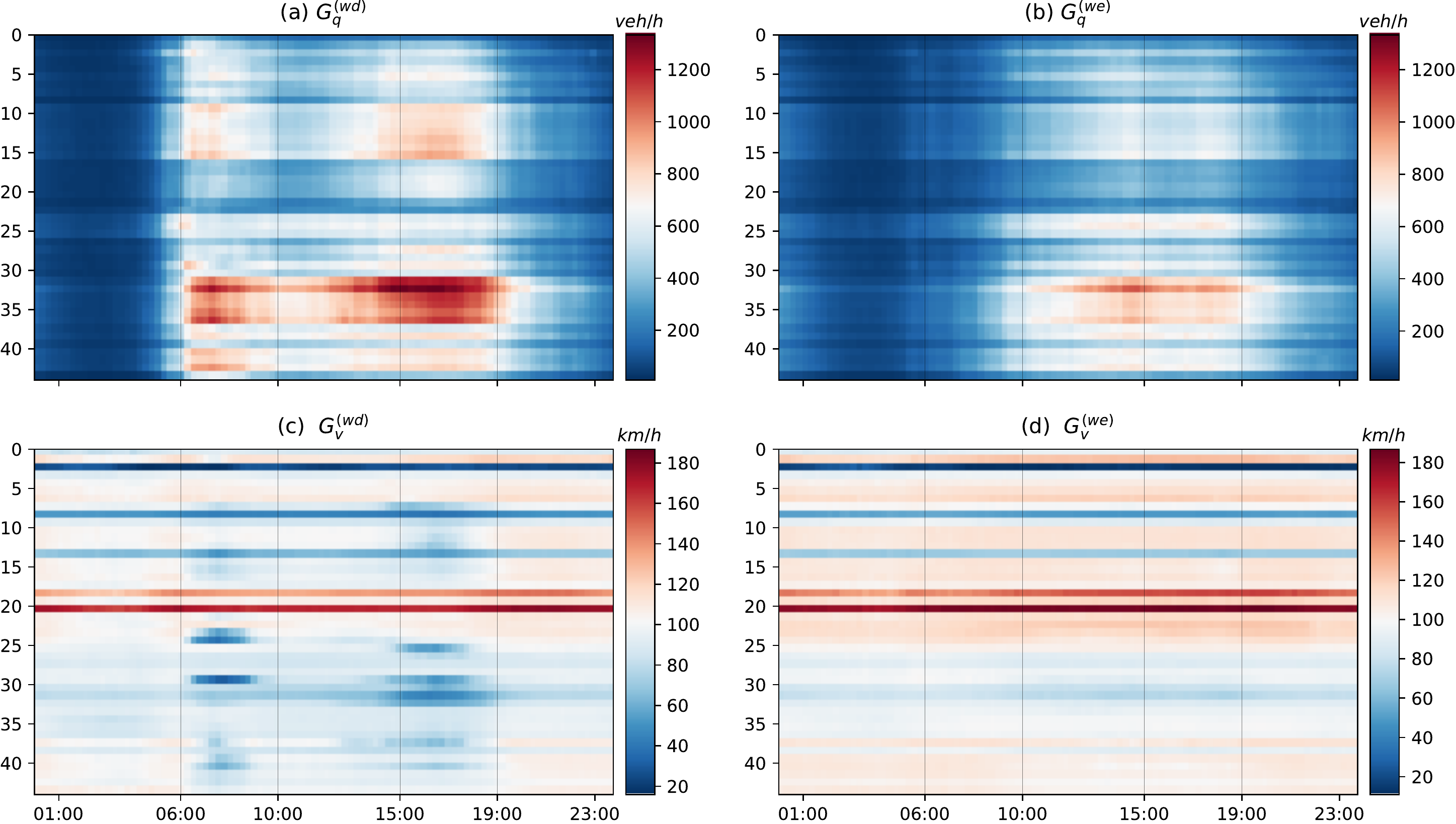}
		\caption{Data matrices $G$ of traffic flow (a)-(b) and velocity (c)-(d) for cross-sections on the motorways A40 and A42 averaged over all workdays ((a) and (c)) and weekends ((b) and (d)), respectively. Vertical black lines separate the different time periods according to Fig. \ref{fig_3_tagesgang}.}
		\label{fig_4_avg_data}
	\end{center}
\end{figure}

\noindent Identifying frequently reoccurring traffic patterns is important to understand correlations of traffic flows or velocities. Therefore an overview of the common traffic situation on the considered motorways is given. In general, the traffic situation on workdays differs from the one on weekends. The overall traffic volume on workdays is significantly higher than on weekends because commuter traffic is present on regular bases, especially in the case of A40. In contrast, the traffic volumes lower on weekends due to the absence of commuter traffic, most notably in the morning. In addition, on Sundays most trucks are not allowed to drive by traffic code until 22:00 in Germany \cite{BuGV}. A representative time evolution of traffic flow on workdays and weekends is shown in Fig.~\ref{fig_3_tagesgang}. The behavior of traffic flow on a typical workday is characterized by peaks during the morning and afternoon rush hours (periods $P_2$ and~$P_4$), while the traffic volume lowers during midday (period $P_3$). In the early evening the traffic flow rate starts to decrease into a minimum (period $P_5$) that holds until the early morning (period $P_1$). A typical weekend day is characterized by a significantly lower overall traffic volume compared to workdays. In the morning (period $P_2$) the traffic flow rate starts to increase, reaching its maximum in the afternoon (during periods $P_3$ and $P_4$) and decreases in the evening again (period $P_5$). The traffic volume during night time (period $P_1$) is higher on weekends than it is on workdays.

To provide an overview of the daily traffic situation on A40 and A42, the data matrices $G$ are considered. A useful representation can be obtained by averaging the data matrices of traffic flows and velocities over all workdays and weekend days, respectively. This simple method has the advantage that frequently occurring traffic patterns (e.g. heavy traffic on regular bases) can be identified, while infrequently occurring traffic patterns (due to accidents, heavy weather conditions etc.) are not standing out. The resulting matrices are shown in Fig.~\ref{fig_4_avg_data}. For a clearer description of the regular traffic situation Tab. \ref{tab:table} summarizes the area of big cities through which the motorways, with the corresponding cross-sections, pass.

The characteristic behavior of traffic flow on workdays (Fig.~\ref{fig_3_tagesgang}) can clearly be recognized in Fig.~\ref{fig_4_avg_data}(a). The averaged data reveals that the largest traffic volume during rush hours commonly occurs around big cities. Motorway A40 is an obvious example, because it connects several big cities which are geographically bordering. Compared to higher traffic flow occurrence around and between Duisburg (cross-sections~24-26) and Mülheim (cross-sections~27-29), the sections connecting Essen with Bochum (cross-sections~31-37) show particularly high average traffic flows during rush hours and in between. This is a well known observation on A40 \cite{Adac_stau}. On A42 the highest average traffic flows during rush hours occur in and around Gelsenkirchen (cross-sections~9-16).

An absence of early morning rush hours and a shift of higher traffic volume to a later time of day in Fig.~\ref{fig_4_avg_data}(b) correspond to the characteristic features for weekends (see Fig.~\ref{fig_3_tagesgang}). As expected, the average traffic flow at most sections is considerably lower compared to workdays. It is noticeable that the sections between Bochum and Essen (sections 31-36) also seem to be the highest trafficked ones on weekends. On the contrary, cross-sections with commonly low traffic volume, e.g. cross-sections 16-21, are easily identifiable.

On workdays, during rush hours, the average velocities at cross-sections with high traffic occurrence are lowered compared to the rest of the day (see Fig.~\ref{fig_4_avg_data}(c)). This may be an indicator for commonly occurring congested traffic, because congestion is more likely to be encountered during time periods of high traffic volumes. Yet, for a precise distinction, a comparison between the average velocities and the minimum velocity for free flow (section \ref{sec32}) is appropriate and will be given in section~\ref{sec42}. In contrast to workdays, the average velocities on weekends do not exhibit any noticeable time periods with conspicuously low values. This may indicate an absence of regular congestion on weekends.

It is noteworthy that a possible speed limitation of $80$ km/h between Mülheim and Bochum (cross-sections~25-37) can clearly be assumed in Fig.~\ref{fig_4_avg_data}. Finally it should be mentioned that the constant and distinct velocities averaged over a whole day for sections 2, 8, 13, 18 and 20 are corrupted due to missing data. The combined velocity for an effective lane may have biases if the data for one or more lanes at a cross-section is missing.

\begin{table}[htbp]
	\caption{An overview of the big cities in NRW through which motorways A40 and A42 with the corresponding cross-sections pass.}
	\begin{center}
		\begin{tabular}{llr}  
			\toprule
			City    & Motorway & Cross-section \\
			\midrule
			Duisburg & A42 & 3      \\
			 & A40 & 24-26       \\
			Oberhausen & A42 & 4-6      \\
			Mülheim & A40 & 27-29      \\
			Essen & A42 & 8-9       \\
			& A40 & 30-33 \\
			Gelsenkirchen & A42 & 10-13 \\
			Bochum & A40 & 35-40 \\
			Dortmund & A40 & 41-43 \\
			\bottomrule
		\end{tabular}		
	\end{center}
	\label{tab:table}
\end{table}

\subsection{Correlation matrices for workdays and weekends}
\label{sec42}
\subsubsection{Pearson correlation matrices and their structural features}
\label{sec421}
\begin{figure}[tbp]
	\begin{center}
		\includegraphics[width=0.6\textwidth]{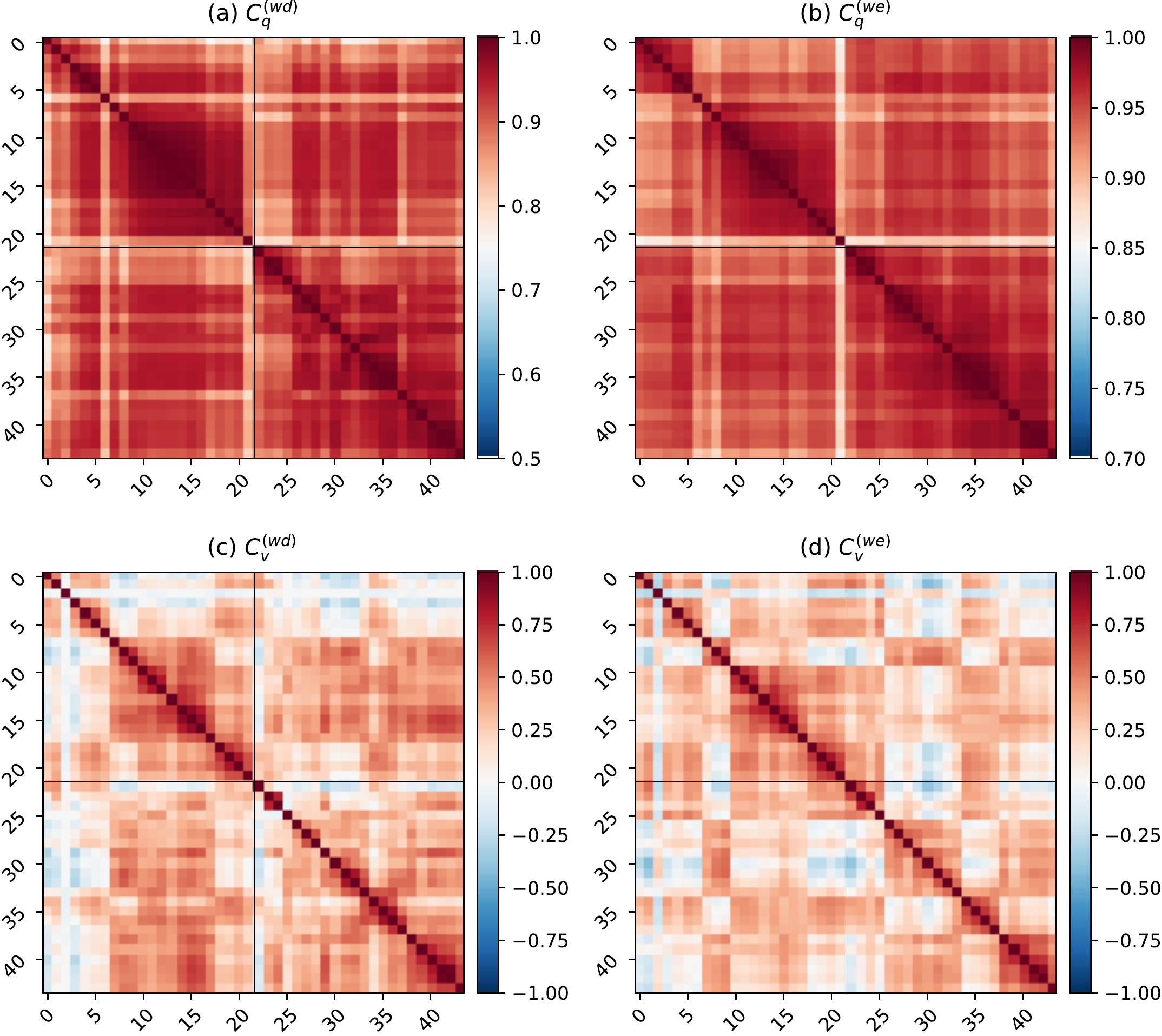}
		\caption{Correlation matrices $C$ of traffic flow (a)-(b) and velocity (c)-(d) for cross-sections on the motorways A40 and A42 averaged over all workdays (wd) and weekends (we). The horizontal and vertical axis indicate the section numbers corresponding to the numbering in Fig. \ref{fig_1_map}. Thin black lines separate the cross-sections of both motorways. For (a) and (b) a different color scale is used to reveal the details of the correlation structure.}
		\label{fig_5_corr_wd_we}
	\end{center}
\end{figure}

Considering the significant differences between workdays and weekends we work out correlation matrices for both cases with Eq. \eqref{eq_corr_mat}. For each day we calculate correlation matrices for traffic flows and velocities. Averaging these matrices over all workdays and weekends, respectively, result in the average correlation matrices. As shown in Fig. \ref{fig_5_corr_wd_we}, the four matrices exhibit distinct patterns, i.e. visually accessible structures. Overall, both correlation matrices of traffic flows show high positive correlations for the sections on A40 (upper left), A42 (lower right) and for both motorways (off-diagonals). The correlation patterns on workdays differ from the ones on weekends. However, in both cases the squared diagonal blocks with high values imply that similar behavior in traffic flows and in velocities is present among sections from both motorways. By averaging over all matrix elements $C_{q,ij}$ we find that the average correlation on weekends $\langle \corr{q,ij}{we} \rangle = 0.9496$ is higher than the one on workdays $\langle \corr{q,ij}{wd} \rangle = 0.9148$. The overall correlation strength of velocities is clearly lower than the one of traffic flows. In contrast to the ones of traffic flows, the correlation matrices of velocities in Fig.~\ref{fig_5_corr_wd_we} manifest more structural features, accompanied with the emergence of negative correlations and more distinct differences between workdays and weekends. In particular, the average correlation for weekends $\langle \corr{q,ij}{we} \rangle = 0.2555$ is lower than the one for workdays $\langle \corr{q,ij}{wd} \rangle=0.3020$, and both values are smaller than the ones regarding traffic flows. Similar to the case of traffic flows, the velocity correlation matrices show diagonal blocks of positive correlation. Some of these blocks are overlapping with the ones for traffic flow. The off-diagonal sub-matrices of $\corr{v}{wd}$ contain a larger rectangular block of positive correlations, which suggest similar behavior of certain sections on A40 and A42. For the sake of completeness we present the distributions of correlation coefficients for all four matrices in Appendix~\ref{ap_distribution}. There we find that the distributions are close enough to a Gaussian distribution to serve the purpose of our work. 

\subsubsection{Comparison of Pearson and Spearman correlation matrices}
\label{sec422}

\begin{figure}[tbp]
	\begin{center}
		\includegraphics[width=0.6\textwidth]{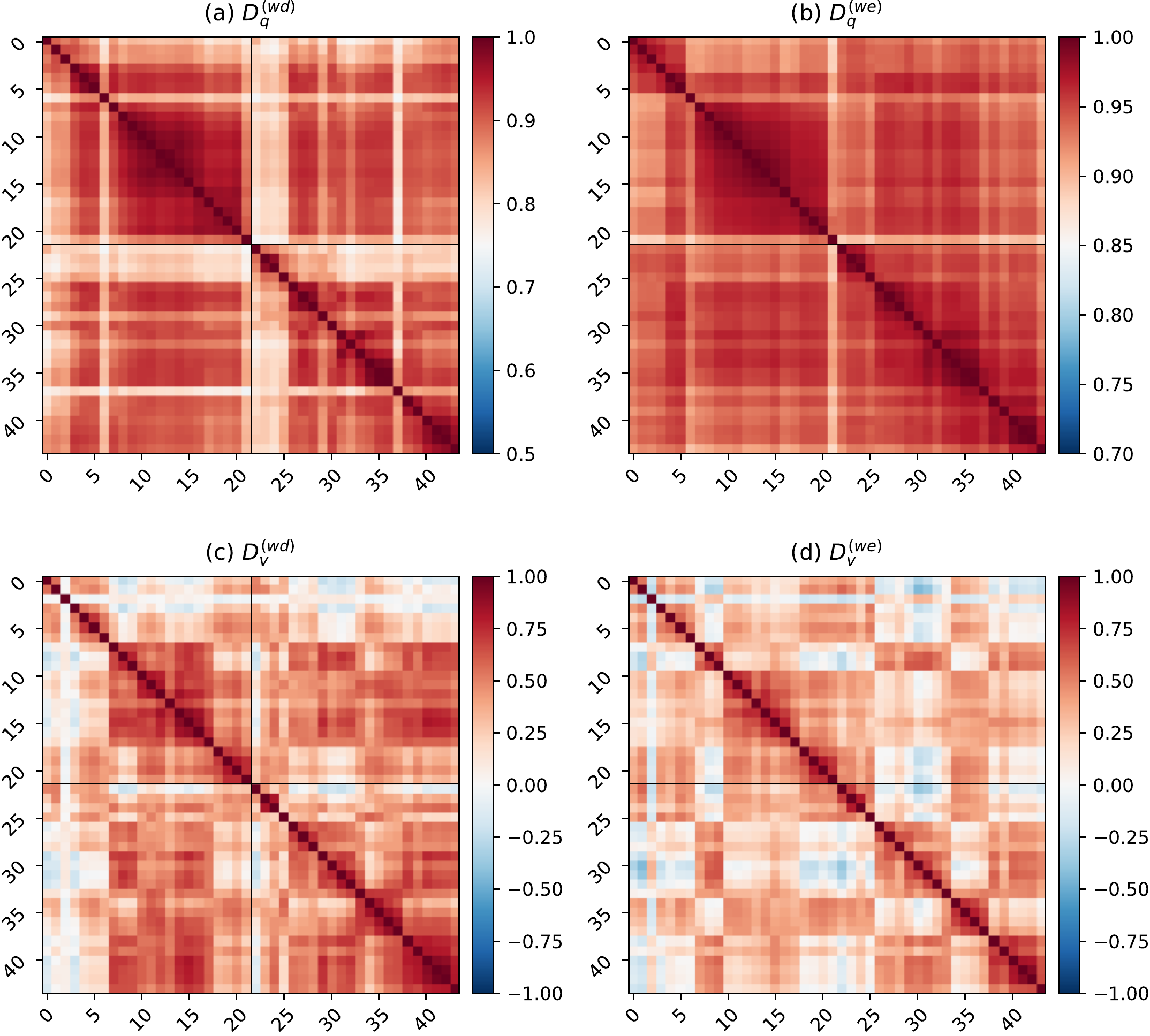}
		\caption{Spearman correlation matrices $D$ of traffic flow (a)-(b) and velocity (c)-(d) for cross-sections on the motorways A40 and A42 averaged over all workdays (wd) and weekends (we). The horizontal and vertical axis indicate the section numbers corresponding to the numbering in Fig. \ref{fig_1_map}. Thin black lines separate the cross-sections of both motorways. For (a) and (b) a different color scale is used to reveal the details of the correlation structure.}
		\label{fig_x_corr_wd_we_sp}
	\end{center}
\end{figure}

For the purpose of our study we calculate the most trivial type of correlation coefficients for the given time series data, Pearson correlation coefficients. Despite, the analysis of structural features in Pearson correlation matrices of time series data is an established method in the field of financial markets \cite{Muennix_2012,Heckens_2020}. Considering the approach to transfer the method of structural analysis of correlation matrices onto traffic data we like to substantiate our results by comparing the Pearson correlations in Fig.~\ref{fig_5_corr_wd_we} with another type of correlation coefficient. Here we chose Spearman correlation. For the calculation of the Spearman correlation coefficients we rank our daily traffic flow and velocity time series (data matrices~$G$, see Eq.~\eqref{eq_data_matrix}) resulting in ranked data matrices~$G_R$ for each day. According to Eq.~\eqref{eq_corr_coef} we subsequently calculate the Pearson correlation coefficients of the ranked data matrices~$G_R$ resulting in Spearman correlation matrices $D$ for each day. Equivalent to the procedure in Section~\ref{sec421} we average the Spearman correlation matrices of traffic flows and velocities over all workdays and weekends, respectively, resulting in averaged Spearman correlation matrices in Fig.~\ref{fig_x_corr_wd_we_sp}.

Comparing the Spearman correlation matrices in Fig.~\ref{fig_x_corr_wd_we_sp} with the Pearson correlation matrices in Fig.~\ref{fig_5_corr_wd_we} reveals remarkable similarities in the correlation patterns. This is the case for both traffic flow and velocity correlations on workdays as well as on weekends. In case of traffic flows both Spearman correlation matrices show overall high positive correlations. Similar to the Pearson correlations, the Spearman traffic flow correlations exhibit different patterns for workdays and weekends. The corresponding squared diagonal block structures in Fig.~\ref{fig_x_corr_wd_we_sp}(a)~and~(b) coincide with the ones found in the Pearson correlation matrices in Fig.~\ref{fig_5_corr_wd_we}(a)~and~(b). Contrary to the traffic flow correlations, the Spearman velocity correlation matrices also manifest more structural features and exhibit an overall lower correlation strength. Nearly all of the block structures found in the Spearman velocity correlation matrices in Fig.~\ref{fig_x_corr_wd_we_sp}(c)~and~(d) coincide with the structures of the Pearson velocity correlation matrices in Fig.~\ref{fig_5_corr_wd_we}(c)~and~(d).

The short comparison above shows the remarkable accordance of the correlation patterns for both types of correlation coefficients. We perform a detailed analysis of the structural features on the basis of the Pearson correlation matrices in the following Section.

\subsubsection{Analysis of structural features in Pearson correlation matrices}
\label{sec423}

Several correlation structures in Fig.~\ref{fig_5_corr_wd_we} are directly in accordance with the information revealed with the data matrices in Fig.~\ref{fig_4_avg_data}. As an example, the coextensive diagonal blocks with strong traffic flow or velocity correlations on workdays among the cross-sections between Essen and Gelsenkirchen (cross-sections~9-16) on A42 correspond to the outstanding time-dependent similarities in the averaged data matrices in Fig.\ref{fig_4_avg_data}. In contrast, the correspondence between the data matrices and the less distinct correlation structures across the cross-sections between Mülheim and Bochum (cross-sections~26-36) on A40 is less obvious. Comparing the lowered average velocities during rush hours on workdays for both cases leads to the conclusion, that congestion occurs more frequently between Mülheim and Bochum on A40 than it does between Essen and Gelsenkirchen on A42. On the one hand this is in accordance with common observations on A40 \cite{Adac_stau}, on the other hand it implies that congested traffic has an influence on the emergence of correlation structures and may be the reason for more indistinct patterns. The latter implication is also supported by the correlation matrices for weekends, because congestion is less likely to occur on weekends due to lower traffic volumes and the structural features regarding the correlations among the cross-sections~26-36 are more distinct compared to workdays. To find evidence, it is necessary to empirically identify free flow and congested traffic states. According to Section~\ref{sec32} we are able to determine the minimum free flow velocity~$v_{\text{min}}^{\text{(free)}}$ for 41 cross-sections by applying k-means clustering to our data. Following this, we determine the proportion of free flow and congested traffic states 
\begin{align}
	n_{k}^{(\text{free})} &= \frac{N_k^{(\text{free})}}{N_k}, \\
	n_{k}^{(\text{cong})} &= \frac{N_k^{(\text{cong})}}{N_k},
\end{align}
for each cross-section, where $N_k^{(\text{free})}$ is the number of free flow states, $N_k^{(\text{cong})}$ the number of congested states and 
\begin{equation}
	N_k = N_k^{(\text{cong})} + N_k^{(\text{free})}
\end{equation}
the total number of states detected at cross-section $k$. Due to inconclusive data the results for determining ~$v_{\text{min}}^{\text{(free)}}$ for cross-sections~2,~13~and~18 are excluded from the further investigation.
\begin{figure}[tbp]
	\begin{center}
		\includegraphics[width=0.8\textwidth]{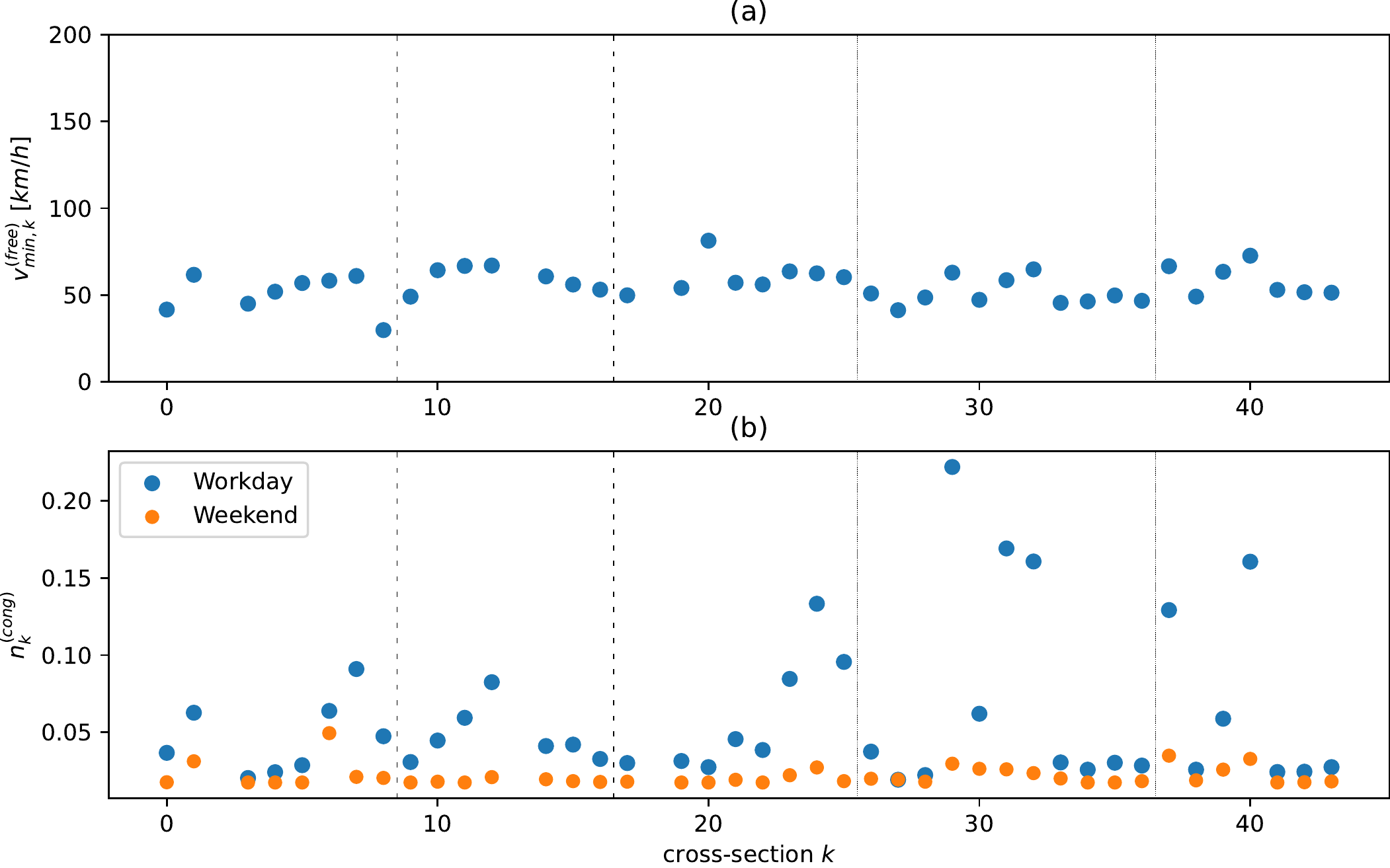}
		\caption{(a) Minimum free flow velocity $v_{\text{min}}^{\text{(free)}}$, (b) proportion of congested traffic states and (c) proportion of free flow traffic states for each cross-section $k$. Vertical black lines indicate cross-sections~9-16 (dashed line) and cross-sections~26-36 (dotted line).} 
		\label{fig_6_v_min_free_and_counts}
	\end{center}
\end{figure}

As shown in Fig.~\ref{fig_6_v_min_free_and_counts}, we find that the minimum free flow velocities for cross-sections~(9-16) between Essen and Gelsenkirchen and cross-sections~(26-36) between Mülheim and Bochum exhibit the same range of values. We also find that the proportions of congestion are in agreement with the hypothesis we derived from Fig.~\ref{fig_4_avg_data}. On workdays most of the cross-sections between Mülheim and Bochum on A40 detected an higher amount of congested states than the cross-sections between Essen and Gelsenkirchen on A42. In addition, the proportions of congested states on weekends show similar values in both cases, whereby most of these values are smaller compared to the ones for workdays. This also fits our expectation of reduced congestion occurrence on weekends due to lower traffic volumes.
\begin{figure}[tbp]
	\begin{center}
		\includegraphics[width=1\textwidth]{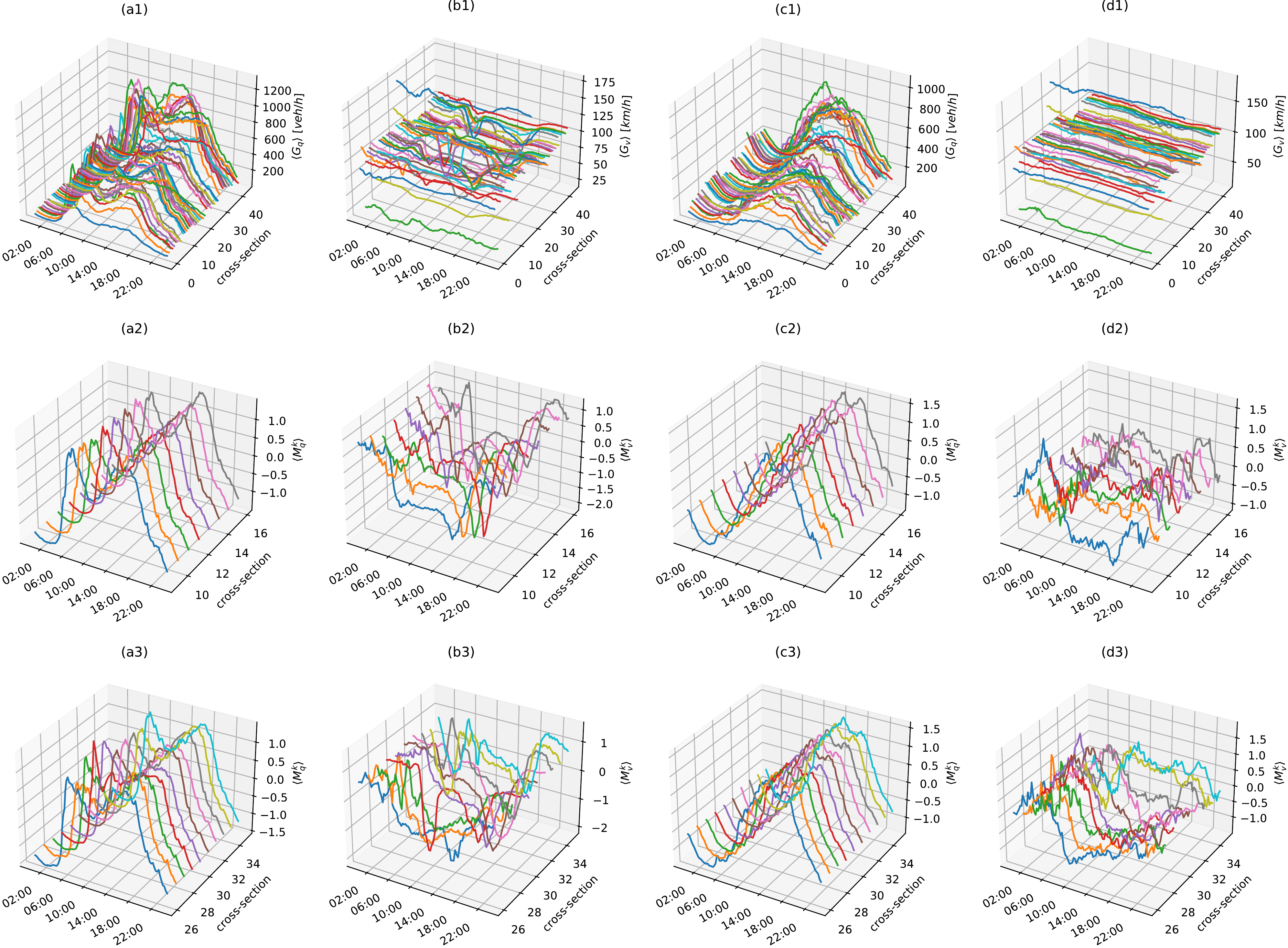}
		\caption{Time evolution of the averaged data $G$ (first row) and samples of the averaged normalized data $M$ (second and third row) for traffic flow (columns (a),(c)) and velocity (columns (b),(d)) for workdays (columns (a),(b)) and weekends (columns (c),(d)) shown in 3-dimensional space.}
		\label{fig_7_avg_norm_data}
	\end{center}
\end{figure}

For a further investigation we consider the underlying normalized data matrices $M$ in Eq.~\eqref{eq:standardize}, which emphasize the similarities in trend between the traffic flow or velocity time series in a clearer way compared to the original data matrices. Similar to Fig.~\ref{fig_4_avg_data} we average the normalized data matrices~$G$ over all workdays and weekends, respectively. As shown in Fig.~\ref{fig_7_avg_norm_data}(a2), (a3), (c2) and (c3), the remarkably high similarities in trend among normalized traffic flows imply that the overall strongly positive correlations of traffic flows in Fig.~\ref{fig_5_corr_wd_we} mainly originate from the daily collective traffic behavior characterized by the time periods in Fig.~\ref{fig_3_tagesgang}. In contrast to the traffic flow data, the more fluctuating nature of the normalized velocities in Fig.~\ref{fig_7_avg_norm_data} (b2), (b3), (d2) and (d3) are the reason for the overall lower correlation strength and the more inhomogeneous structures in Fig.~\ref{fig_5_corr_wd_we}. 

Comparing the averaged normalized traffic flows or velocities of the highly correlated cross-sections between Essen and Gelsenkrichen on A42 (Fig.~\ref{fig_7_avg_norm_data}(a2)-(b2) and (c2)-(d2)) with the ones of lower average correlation between Mülheim and Bochum on A40 (Fig.~\ref{fig_7_avg_norm_data}(a3)-(b3) and (c3)-(d3), respectively) points out the differences in temporal evolution among the averaged time series. These differences lead to the more indistinct structural features in the average correlation matrices in Fig.~\ref{fig_5_corr_wd_we}. Further, the differences mostly emerge during the rush hours which we consider as further evidence for the impact of congested traffic on the structural differences in the correlation matrices. 
 \begin{figure}[tbp]
	\begin{center}
		\includegraphics[width=0.85\textwidth]{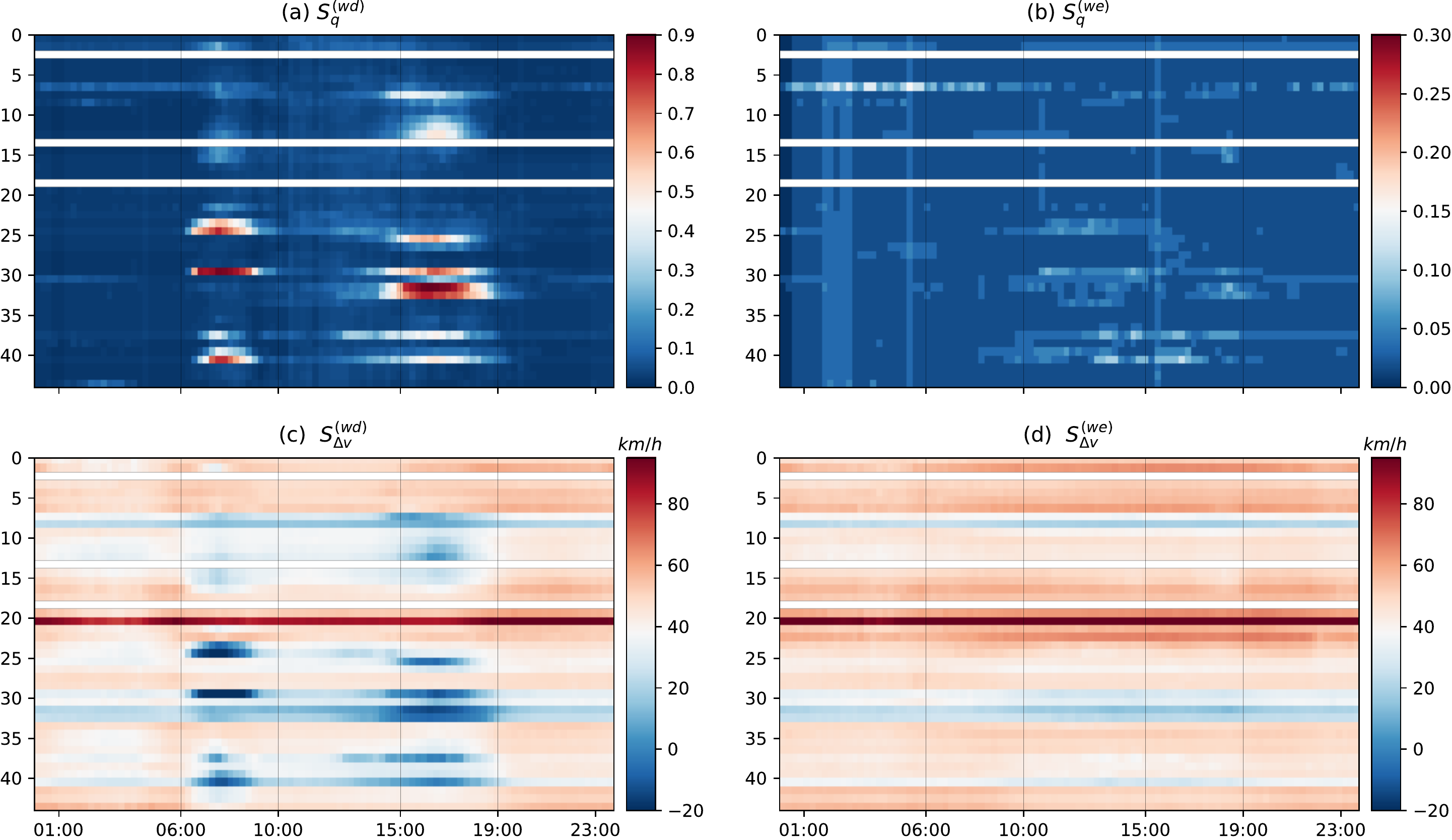}
		\caption{Matrices $S_{q}$ (a)-(b) and $S_{\Delta v}$ (c)-(d) averaged over all workdays (wd) and weekends (we). The matrices $S_{q}$ contain the probability to find a congested state at a time $t$ at cross-section $k$ in the studied data. The averaged matrices $S_{\Delta v}$ contain the averaged velocity differences $\Delta v_k (t)$. Positive (negative) values for $\Delta v_k (t)$ indicate free flow (congested) traffic states.}
		\label{fig_8_state_mapping}
	\end{center}
\end{figure}

Comparing information about the distribution of traffic states in space and time with the correlation matrices in Fig.~\ref{fig_5_corr_wd_we} may demonstrate the influence of congestion on the correlation structures in a clear fashion. To this end, we work out an average time evolution of traffic states for all cross-sections. Therefore we map the identified traffic states onto a new set of time series. For each day and cross-section~$k$ we calculate
\begin{equation}
	S_{k}^{(q)}(t) = 
	\begin{cases}
		1 & \text{if } \frac{q_k(t)}{\rho_k(t)} < v_{\text{min,k}}^{(\text{free})}, \\
		0 & \text{if } \frac{q_k(t)}{\rho_k(t)} \geq v_{\text{min,k}}^{(\text{free})},
	\end{cases}
\label{eq_mapping_1}
\end{equation}
and
\begin{equation}
	S_{k}^{(\Delta v)}(t) = v_k(t)-v_{\text{min,k}}^{(\text{free})},
	\label{eq_mapping_2}
\end{equation}
for workdays and weekends separately. These time series are the rows of the matrices $S_{q}$ and $S_{\Delta v}$, which we average over all workdays and weekends, respectively. We consider both matrices as an average traffic state of the studied motorways. The matrix~$S_{q}$ contains the probabilities to find a congested traffic state at cross-section $k$ at a given time $t$ within our data. The matrix~$S_{\Delta v}$ contains the averaged velocity difference $\Delta v_k = v_k(t)-v_{\text{min},k}^{(\text{free})}$ which indicates either free-flow ($\Delta v_k > 0$) or congestion ($\Delta v_k < 0$), whereas its magnitude $|\Delta v_k|$ indicates the degree of a respective state \cite{Wang2020}. Combining the information of both matrices should provide a sufficient overview over the distribution of traffic states. 

We find that the similarities and differences among the traffic state evolution of pairwise cross-sections in Fig.~\ref{fig_8_state_mapping} provide a good explanation for most structural features in the correlation matrices in Fig.~\ref{fig_5_corr_wd_we}. As a first example we consider cross-sections~9-16 between Essen and Gelsenkirchen on A42. On workdays congestion mostly occurs during the afternoon rush hours while free flow states govern the rest of a day. The congested states at all cross-sections~(9-16) begin to emerge around the same time in the afternoon (Fig.~\ref{fig_8_state_mapping}(a)) and exhibit with a higher degree than the occasional congestion during the morning rush hours (Fig.~\ref{fig_8_state_mapping}(b)). These similarities in time evolution of traffic states result in high time-dependent similarities among the original time series, which provide a reasonable explanation for the corresponding strongly correlated blocks in Fig.~\ref{fig_5_corr_wd_we}(a) and (b). 

In contrast to this example, we examine cross-sections~26-36 between Mülheim and Bochum on A40, which show a very different evolution of traffic states on workdays. While some cross-sections~(23-25,29) exhibit congestion during the morning rush hours, other cross-sections~(29-34) manifest severe congestion during the afternoon rush hours. Despite the probability to encounter occasional congestion during the afternoon at certain cross-sections~(26-28) in Fig.~\ref{fig_8_state_mapping}(a) the corresponding velocity differences in Fig.~\ref{fig_8_state_mapping}(b), as well as the proportions in Fig.~\ref{fig_6_v_min_free_and_counts}(b) and (c), suggest that free-flow is the more dominating traffic state during these time periods. We conclude that the strongly differing distributions of congestion and free flow among cross-sections~26-36 in Fig.~\ref{fig_8_state_mapping} result in less time-dependent similarities among the original traffic flow time series which lead to the corresponding indistinct structural features in the correlation matrices in Fig.~\ref{fig_5_corr_wd_we}(a) and (b).

The examples above demonstrate that the correlation matrices are able to capture the average traffic state in Fig.~\ref{fig_8_state_mapping}. This finding is further supported by taking into account the matrices for weekends. A pairwise comparison shows that certain block structures of the traffic flow correlations in Fig.~\ref{fig_5_corr_wd_we}(b) are in good agreement with the traffic state evolution in Fig.~\ref{fig_8_state_mapping}(b) (e.g. cross-sections~33-36~and~40-42). Despite occasional congestion, the strongly dominating free flow states are the reason for the emergence of more distinct block structures, which becomes particularly clear in case of the velocity correlations in Fig.~\ref{fig_8_state_mapping}(d). These block structures correspond well to the average traffic state in Fig.~\ref{fig_8_state_mapping}(d) (e.g., cross-sections~3-6,9-12,~26-29,~33-36~and~38-44).
 
\subsection{Correlation matrices for individual time periods}

The correlation matrices in Fig.~\ref{fig_5_corr_wd_we} are able to reflect the average traffic state in Fig.~\ref{fig_8_state_mapping}. The dominance of free flow and congested states varies depending on the individual time periods, therefore we expect that correlation matrices of individual time periods contribute different structural features to the correlation matrices regarding the whole day. To gain an impression of these contributions, we work out the correlation matrices for the five individual time periods shown in Fig.~\ref{fig_3_tagesgang}. 
\begin{figure}[tbp]
	\begin{center}
		\includegraphics[width=1\textwidth]{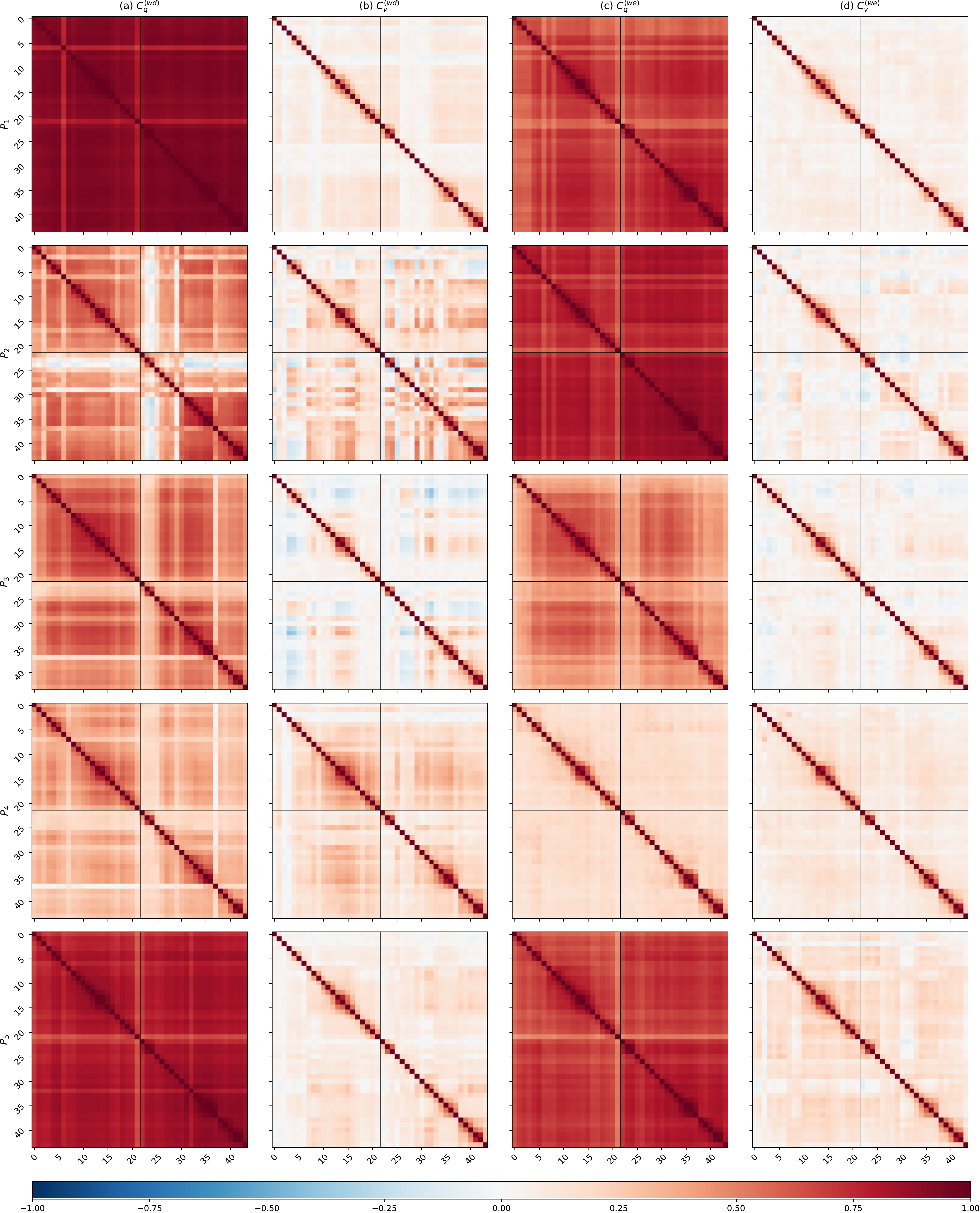}
		\caption{Overview of the traffic flow (columns (a) and (c)) and velocity (columns (b) and (d)) correlation matrices of the individual time periods $P_1$-$P_5$ for workdays (columns (a) and (b)) and weekends (columns (c) and (d)) according to Fig.~\ref{fig_3_tagesgang}. Each row contains the matrices for a time period in ascending order from $P_1$ to $P_5$.}
		\label{fig_9_corr_periods}
	\end{center}
\end{figure}

In contrast to the case for a whole day, the individual time periods have a shorter length of time leading to less available data points for the calculation of the correlation matrices. To obtain a sufficient amount of data points, we aggregate our data to time intervals of 3 minutes. Therefore the length of the time series in the data matrices~$G$ change to $T$, where $T \in \{80,100,120\}$. For each time period we calculate the daily correlation matrices which are averaged over all workdays and weekends similar to the case in section \ref{sec41}.

As shown in Fig.~\ref{fig_9_corr_periods}, we find that time periods of low traffic volume, namely $P_1$ and $P_5$ on workdays as well as $P_1$, $P_2$ and $P_5$ on weekends, show striking high positive traffic flow correlations. Considering the information gathered from Figures~\ref{fig_4_avg_data} and \ref{fig_8_state_mapping} it is reasonable to assume that the dominance of free flow states during these time periods is the driving force for the overall high strength in correlations. In contrast to this, the correlation matrices for time periods with heavy traffic ($P_2$-$P_4$ on workdays and $P_3$-$P_4$ on weekends) exhibit more distinct structural features and an overall lower level of correlation strength in general. In view of the results revealed from Fig.~\ref{fig_8_state_mapping}, it is reasonable to assume that the lower strength in correlation results from the emergence of congested traffic states. 

Most of the occurring structures in the correlation matrices in Fig.~\ref{fig_9_corr_periods} (column (a) and (c)) clearly coincide with structural features found in the correlation matrices for a whole day in Fig.~\ref{fig_5_corr_wd_we}, e.g. diagonal blocks regarding cross-sections 7-16, 26-36 and 30-42. A corresponding comparison reveals that some of these structures are in remarkable agreement with the average traffic states in Fig.~\ref{fig_8_state_mapping}. This is well demonstrated by considering the block structure of high correlation regarding cross-sections~31-36 in the morning rush hours ($P_2$) on workdays. The corresponding traffic state evolution at these cross-sections in Fig.~\ref{fig_8_state_mapping}(a) clearly show high time-dependent similarities due to a low probability to encounter congestion. In the case of weekends, the revealed coincidences of the structural features are less obvious, but they are present among the traffic flow correlation matrices.
\begin{figure}[tbp]
	\begin{center}
		\includegraphics[width=0.8\textwidth]{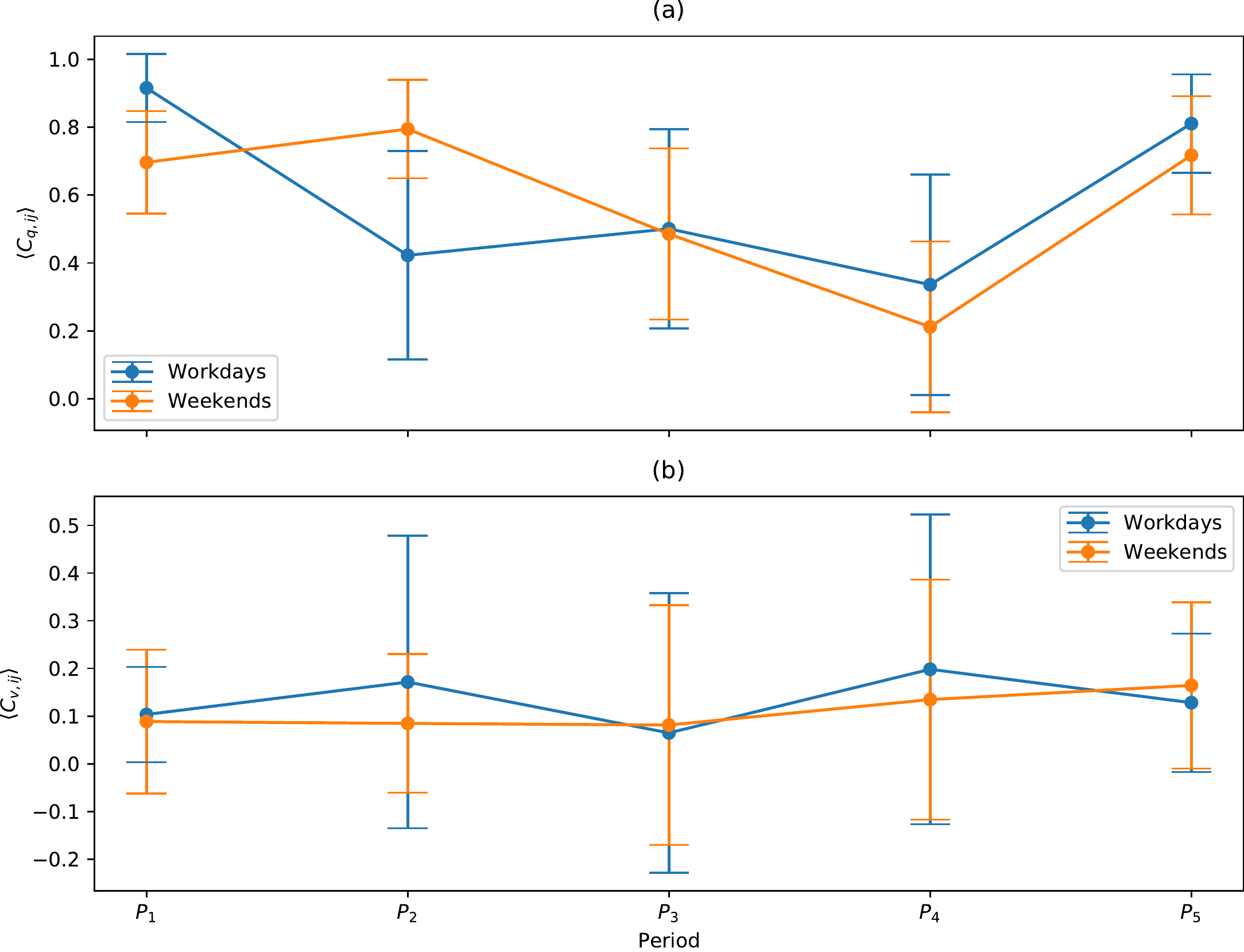}
		\caption{Average correlation of traffic flow (a) and velocity (b) for the different time periods on workdays and weekends. Error bars indicate the standard deviation.}
		\label{fig_10_avg_corr}
	\end{center}
\end{figure}

The above findings strongly suggest that the correlation matrices of the individual time periods contribute different structural features to the correlation matrices regarding the whole day in Fig.~\ref{fig_5_corr_wd_we}, as initially expected. This is also the case for the velocity correlation matrices of the different time periods. Comparing the structural features in the correlation matrices for velocities during periods~$P_2$-$P_4$ in Fig.~\ref{fig_9_corr_periods} (columns (b) and (d)) with the correlation matrices in Fig.~\ref{fig_5_corr_wd_we} demonstrates this particularly clear. In view of the coinciding correlation structures for both, traffic flow and velocity correlations, one may be tempted to derive the hypothesis that the correlation matrices for the whole day in Fig.~\ref{fig_5_corr_wd_we} can be interpreted as a superposition of the correlation matrices for the individual time periods in Fig.~\ref{fig_9_corr_periods}.

The different overall strength in the correlation matrices for the individual time periods in Fig.~\ref{fig_9_corr_periods} suggest an opposite behavior of traffic flow and velocity correlations. While the overall correlation strength of traffic flow is decreased during time periods of heavy traffic, the velocity correlation strength increases for workdays~($P_2$-$P_4$) as well as for weekends~($P_3$-$P_4$). To gain an impression of how strong this effect is, we calculate the traffic flow and velocity correlations~$\langle C_{q,ij} \rangle$~and~$\langle C_{v,ij} \rangle$ averaged over all elements in their corresponding matrices.

As shown in Fig.~\ref{fig_10_avg_corr}, the average correlation of traffic flows and velocities, and the corresponding standard deviations, show the presumed opposite behavior and are able to capture time periods of high and low traffic volumes. The average correlation of traffic flows on workdays and weekends is clearly lower for the time periods with heavy traffic ($P_2$-$P_4$ on workdays and $P_3$-$P_4$ on weekends). Also, the average correlation is able to reflect the increased traffic volume on weekends during the night compared to the morning (see. Fig.~\ref{fig_3_tagesgang} and Fig.~\ref{fig_7_avg_norm_data}(c1)). In contrast, the average correlation of velocities exhibit increased values for the rush hours on workdays as well as for the afternoon and night on weekends. Since the correlation matrices are able to capture the average traffic state, it is reasonable to assume that the average correlation is able to reflect the proportions of congestion and free flow. Considering the average traffic states in Fig.~\ref{fig_8_state_mapping} and the results from Fig.~\ref{fig_10_avg_corr} this is very likely to be the case.

A possible explanation for the behavior of average correlation in Fig.~\ref{fig_10_avg_corr} is the time evolution of traffic states during the individual time periods. In the evening and night the traffic volumes are low at all cross-sections and free flow states dominate, which leads to an high average correlation. During rush hours the probability to encounter congestion is increased at several cross-sections (see Fig.~\ref{fig_8_state_mapping}) which results in the emergence of both states at different times and cross-sections. As a result, the traffic flows experience fluctuations which lower the degree of similarity among them and therefore the average correlation. In case of velocity it is reasonable to assume the opposite. During time periods of high traffic occurrence, especially when encountering congestion, the traffic participants are forced to adapt their velocities due to vehicle interactions. Therefore we expect the velocities to experience less fluctuations during rush-hours and in between, resulting in an higher average correlation. 

\section{Conclusions}
\label{sec5}

Correlation structures of traffic flow and velocity time series, observed on the German motorways A40 and A42, were analyzed. Due to the differences in time evolution of traffic flow and the overall amount of traffic volume, we distinguished the correlations on workdays and weekends. The comparison of correlation matrices and averaged data suggests that the correlation structures are able to reflect the emergence of frequent congestion at certain cross-section. To demonstrate this, we determined the proportions of congestion and free flow states within our data. The proportions of congested traffic states reveal that certain cross-sections, which show more irregular block structures within the corresponding part of the correlation matrices, exhibit a higher probability for congestion. 

The investigation of averaged normalized traffic flows reveals remarkable similarities in trend, which suggest that the overall strongly positive traffic flow correlations mainly originate from the daily collective traffic behavior. In terms of normalized traffic flows and velocities, we compared groups of cross-sections with higher and lower correlation. We found that differences in time evolution of both quantities mainly emerge during time periods of high traffic volume, which we consider further evidence for the reflection of congestion through the correlation structures.

Motivated by our findings we mapped traffic states onto time series for each cross-section, which resulted in an average traffic state of the studied motorways. The comparison between traffic flow or velocity correlation matrices and the traffic state evolution at several cross-sections revealed that the structural features of the correlation matrices are able to capture and reflect the averaged traffic state we derived from our data. Therefore, the conducted correlation analysis provides useful indicators for critical parts of the road network in terms of frequently congested motorway segments and environmental factors, e.g. constructions sites.

Since the strength in dominance of free flow and congestion depends on time (e.g. rush hours) we calculated the correlation matrices of traffic flows and velocities for individual time periods. We revealed that the correlation matrix for each time period contributes different structural features to the correlation matrix regarding a whole day. The comparison of correlation matrices for individual time periods with the average traffic state demonstrated that several structural features correspond very well with the time dependent state evolution of the studied cross-sections.

Considering the different overall strength in correlations among the time periods we calculated the corresponding average correlation for traffic flows and velocities. We discovered that the average correlation is able to capture the proportion of free flow and congestion during the individual time periods. Therefore time periods with high (low) traffic volumes exhibit low (high) values for the average traffic flow correlation. In case of the average velocity correlation we found this behavior inverted. 

\section*{Acknowledgements}
We gratefully acknowledge funding via the grant “Korrelationen und deren Dynamik in Autobahnnetzen”, Deutsche Forschungsgemeinschaft (DFG, 418382724). We thank Straßen.NRW for providing the empirical traffic data. We also thank Die Autobahn GmbH des Bundes (formerly Straßen.NRW) for providing data on construction sites on motorway A40.

\section*{Author contributions}
M.S. and T.G. proposed the research. S.G. prepared the traffic data, performed all calculations, and wrote the manuscript with input from S.W., T.G. and M.S.. All authors contributed equally to analyzing the results and reviewing the paper.

\bibliographystyle{unsrturl}

\begin{thebibliography}{50}
	\bibitem{UmBund} German Federal Environment Agency. \url{https://www.umweltbundesamt.de/bild/weltweiter-autobestand}. Accessed: 2021-09-01.
	\bibitem{Ladyman2013} J. Ladyman, J. Lambert, and K. Wiesner. What is a complex system? \emph{Eur. J. Philos. Sci.},
	3(1):33-67, 2013. doi: 10.1007/s13194-012-0056-8.
	\bibitem{Treiber2010} M. Treiber and A. Kesting. \emph{Verkehrsdynamik und -simulation}. Springer Berlin Heidelberg, Berlin,
	Heidelberg, 2010. doi:10.1007/978-3-642-05228-6\_1.
	\bibitem{Kerner_2009} B. S. Kerner. \emph{Introduction to Modern Traffic Flow Theory and Control: The Long Road to Three-Phase Traffic Theory}. Springer, 2009. doi: 10.1007/978-3-642-02605-8.
	\bibitem{Neubert_1999} L. Neubert, L. Santen, A. Schadschneider, and M. Schreckenberg. Single-vehicle data of highway traffic: A statistical analysis. \emph{Phys. Rev. E}, 60:6480-6490, 1999. doi: 10.1103/PhysRevE.60.6480.
	\bibitem{Tilch_2000} B. Tilch and D. Helbing. Evaluation of single vehicle data in dependence of the vehicle-type, lane, and site. In Dirk Helbing, Hans J. Herrmann, Michael Schreckenberg, and Dietrich E. Wolf, editors, \emph{Traffic and Granular Flow '99}, pages 333-338, Berlin, Heidelberg, 2000. Springer Berlin Heidelberg. doi: 10.1007/978-3-642-59751-0\_31.
	\bibitem{Lee_2000} H. Y. Lee, H.-W. Lee, and D. Kim. Phase diagram of congested traffic flow: An empirical study. \emph{Phys. Rev. E}, 62:4737-4741, 2000. doi: 10.1103/PhysRevE.62.4737.
	\bibitem{Kerner_2002_E} B. S. Kerner. Empirical macroscopic features of spatial-temporal traffic patterns at highway bottlenecks. \emph{Phys. Rev. E}, 65:046138, 2002. doi: 10.1103/PhysRevE.65.046138.
	\bibitem{Schoernhof_2004} M. Schoenhof and D. Helbing. Empirical features of congested traffic states and their implications for traffic modeling. \emph{Transp. Sci.}, 41, 2004. doi: 10.1287/trsc.1070.0192
	\bibitem{Bertini_2005} R. Bertini and M. Leal. Empirical study of traffic features at a freeway lane drop. \emph{J. Transp. Eng.}, 131, 2005. doi: 10.1061/(ASCE)0733-947X(2005)131:6(397).
	\bibitem{NaSch_1992} K. Nagel and M. Schreckenberg. A cellular automaton model for freeway traffic. J. Phys. I, 2:2221, 1992. doi:10.1051/jp1:1992277.
	\bibitem{Schadschneider_1993} A. Schadschneider and M. Schreckenberg. Cellular automation models and traffic flow. \emph{J. Phys. A: Math. Gen.}, 26(15):L679-L683, 1993. doi: 10.1088/0305-4470/26/15/011.
	\bibitem{Schreckenberg_1995} M. Schreckenberg, A. Schadschneider, K. Nagel, and N. Ito. Discrete stochastic models for traffic flow. \emph{Phys. Rev. E}, 51:2939-2949, 1995. doi: 10.1103/PhysRevE.51.2939.
	\bibitem{Barlovic_1998} R. Barlovic, L. Santen, A. Schadschneider, and M. Schreckenberg. Metastable states in cellular automata for traffic flow. \emph{Eur. Phys. J. B}, 5(3):793-800, 1998. doi: 10.1007/s100510050504.
	\bibitem{Kerner_2002} B. S. Kerner, S. L. Klenov, and D. E. Wolf. Cellular automata approach to three-phase traffic theory. \emph{J. Phys. A: Math. Gen.}, 35(47):9971-10013, 2002. doi: 10.1088/0305-4470/35/47/303.
	\bibitem{Knospe_2002} W. Knospe, L. Santen, A. Schadschneider, and M. Schreckenberg. A realistic two-lane traffic model for highway traffic. \emph{J. Phys. A: Math. Gen.}, 35(15):3369-3388, 2002. doi: 10.1088/0305-4470/35/15/302.
	\bibitem{Schadschneider_2010} A. Schadschneider, D. Chowdhury, and K. Nishinari. \emph{Stochastic Transport in Complex Systems: From Molecules to Vehicles}. Elsevier Science, 2010. doi: 10.1016/B978-0-444-52853-7.00016-6.
	\bibitem{Muennix_2012} M. C. Münnix, T. Shimada, R. Schäfer, F. Leyvraz, T. H. Seligman, T. Guhr, and H. E. Stanley. Identifying states of a financial market. \emph{Sci. Rep.}, 2(1):644, 2012. doi: 10.1038/srep00644.
	\bibitem{Chetalova_2015} D. Chetalova, R. Schäfer, and T. Guhr. Zooming into market states. \emph{J. Stat. Mech. Theory Exp.},
	2015(1):P01029, 2015. doi: 10.1088/1742-5468/2015/01/p01029.
	\bibitem{Rinn_2015} P. Rinn, Y. Stepanov, J. Peinke, T. Guhr, and R. Schäfer. Dynamics of quasi-stationary systems: Finance as an example. \emph{Europhys. Lett. (EPL)}, 110(6):68003, 2015. doi: 10.1209/0295-5075/110/68003.
	\bibitem{Stepanov_2015} Y. Stepanov, P. Rinn, T. Guhr, J. Peinke, and R. Schäfer. Stability and hierarchy of quasistationary states: financial markets as an example. \emph{J. Stat. Mech. Theory Exp.}, 2015(8):P08011, 2015. doi: 10.1088/1742-5468/2015/08/p08011.
	\bibitem{Heckens_2020} A. J. Heckens, S. M. Krause, and T. Guhr. Uncovering the dynamics of correlation structures relative to the collective market motion. \emph{J. Stat. Mech. Theory Exp.}, 2020(10):103402, 2020. doi: 10.1088/1742-5468/abb6e2.
	\bibitem{Wang2020} S. Wang, S. Gartzke, M. Schreckenberg, and T. Guhr. Quasi-stationary states in temporal correlations for traffic systems: Cologne orbital motorway as an example. \emph{J. Stat. Mech. Theory Exp.}, 2020(10):103404, 2020. doi: 10.1088/1742-5468/abbcd3.
	\bibitem{Eurostat} Eurostat. Statistics on European Cities. \url{https://ec.europa.eu/eurostat/statistics-explained/index.php?title=Statistics_on_European_cities}. Accessed: 2021-09-01.
	\bibitem{Adac_stau} Allgemeiner Deutscher Automobil-Club (ADAC). Staubilanz. \url{https://www.adac.de/-/media/pdf/vek/fachinformationen/statistiken/staubilanz-adac-statistik.pdf}. Accessed: 2021-09-01.
	\bibitem{OSML} Copyright and License for OpenStreetMap. \url{https://www.openstreetmap.org/copyright}.
	\bibitem{OpDbLi} Open Data Commons Open Database License v1.0. \url{https://opendatacommons.org/licenses/odbl/1-0/}.
	\bibitem{StaDes} Map tiles by Stamen Design licensed under CC BY 3.0. \url{http://creativecommons.org/licenses/by/3.0}, \url{https://stamen.com/}.
	\bibitem{Li_2017} L. Li and X. Chen. Vehicle headway modeling and its inferences in macroscopic/microscopic traffic flow theory: A survey. Transp. Res. Part C Emerg. Technol., 76:170-188, 2017. doi:~10.1016/j.trc.2017.01.007.
	\bibitem{Krbalek_2016} M. Krbalek and T. Hobza. Inner structure of vehicular ensembles and random matrix theory. Phys. Lett. A, 380, 08 2015. doi:~10.1016/j.physleta.2016.03.037.
	\bibitem{Kianfar_2013} J. Kianfar and P. Edara. A data mining approach to creating fundamental traffic flow diagram. Procedia Soc. Behav. Sci., 104:430-439, 2013. doi: 10.1016/j.sbspro.2013.11.136.
	\bibitem{Kianfar_2010} J. Kianfar and P. Edara. Optimizing freeway traffic sensor locations by clustering global-positioning-system-derived speed patterns. IEEE Trans. Intell. Transp. Syst., 11(3):738-747, 2010. doi:10.1109/TITS.2010.2051329.
	\bibitem{BuGV} German Federal Ministry of Justice and Consumer Protection. German Federal Office of Justice. \url{https://www.gesetze-im-internet.de/stvo_2013/__30.html}. Accessed: 2021-09-01.
	\bibitem{construction_2017} Die Autobahn GmbH des Bundes (formerly Straßen.NRW). Data on construction sites on motorway A40 from 2017.
\end{thebibliography}

\addcontentsline{toc}{section}{References}
\clearpage

\begin{appendices}
\section{}
\subsection{Example of Unprocessed Traffic Data}
\label{ap_data_ex}

\begin{figure}[htbp]
	\begin{center}
		\includegraphics[width=0.7\textwidth]{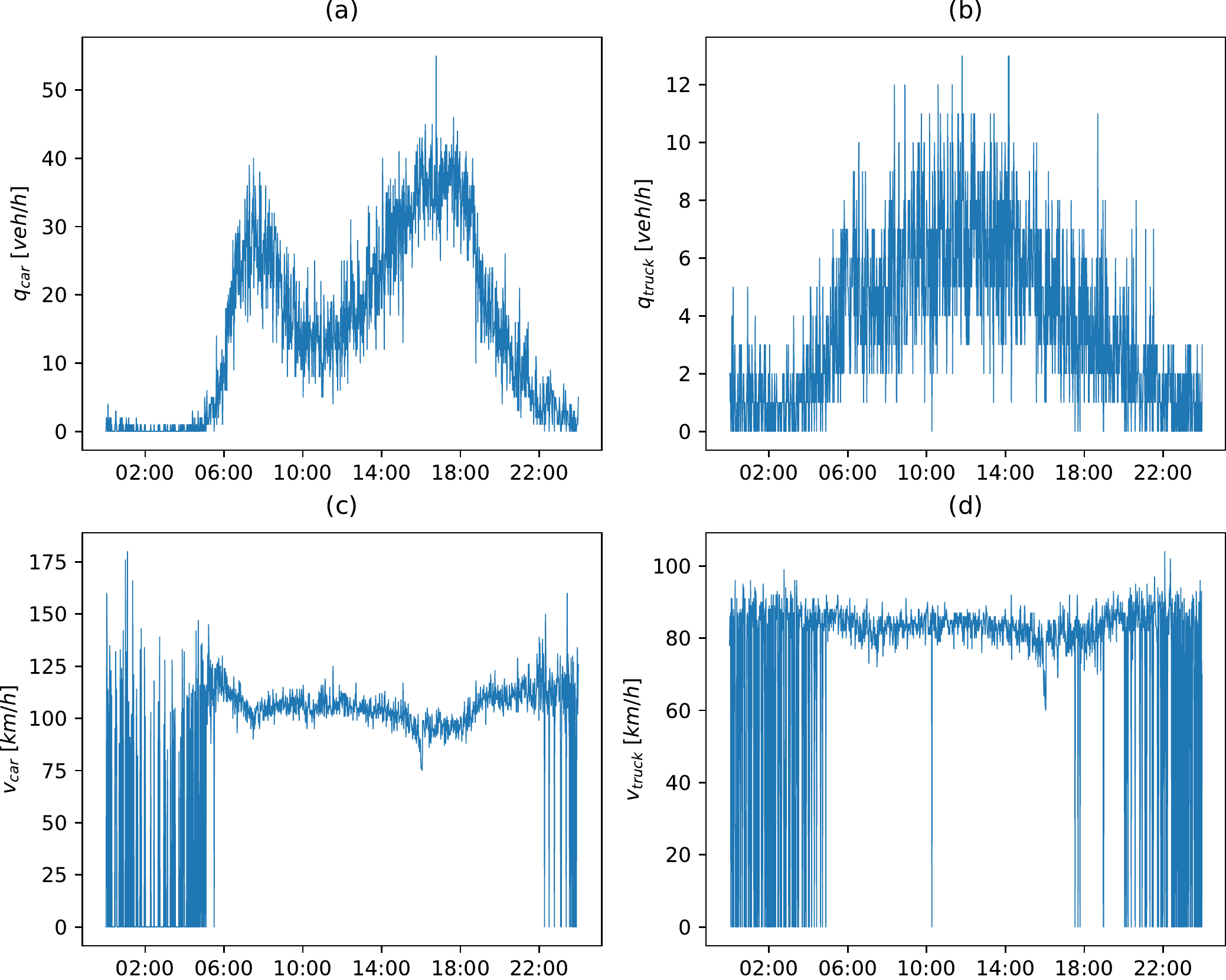}
		\caption{Example of unprocessed traffic flow ((a) and (b)) and average velocity ((c) and (d)) time series data of cars ((a) and (c)) and trucks ((b) and (d)) gathered by the stationary loop detectors at cross-section 33 on motorway A40 on Wednesday, September 20, 2017. The data for cars was detected on the most left lane while the data for trucks was detected on the most right lane.}
		\label{fig_11_data_ex}
	\end{center}
\end{figure}

For the sake of completeness we give an example of the unprocessed time series data used for conducting our study. Figure~\ref{fig_11_data_ex} contains two examples of time series for each quantity (traffic flow $q$ and average velocity $v$) gathered by the stationary loop detectors at cross-section 33 on motorway A40 on Wednesday, September 20, 2017. Figure~\ref{fig_11_data_ex}(a) and (c) show the traffic flow and average velocity time series of cars detected on the most left lane. The typical shape of a traffic flow time series on workdays, including rush hours and non-rush hours, can clearly be identified in Fig.~\ref{fig_11_data_ex}(a) (compare with the processed time series in Fig.~\ref{fig_3_tagesgang} in Section \ref{sec41}). Figure \ref{fig_11_data_ex}(b) and (d) show the traffic flow and average velocity time series of trucks detected on the most right lane. The traffic flow of trucks has its minimum during night time, starts to increase during morning rush hours and reaches its maximum before the afternoon rush hours. The limited speed capability of trucks can be observed in the average velocity time series in Fig.~\ref{fig_11_data_ex}(d).

\clearpage
\subsection{Average Correlation Matrices with shorter Aggregation Time}
\label{ap_3min}

\begin{figure}[htbp]
	\begin{center}
		\includegraphics[width=0.7\textwidth]{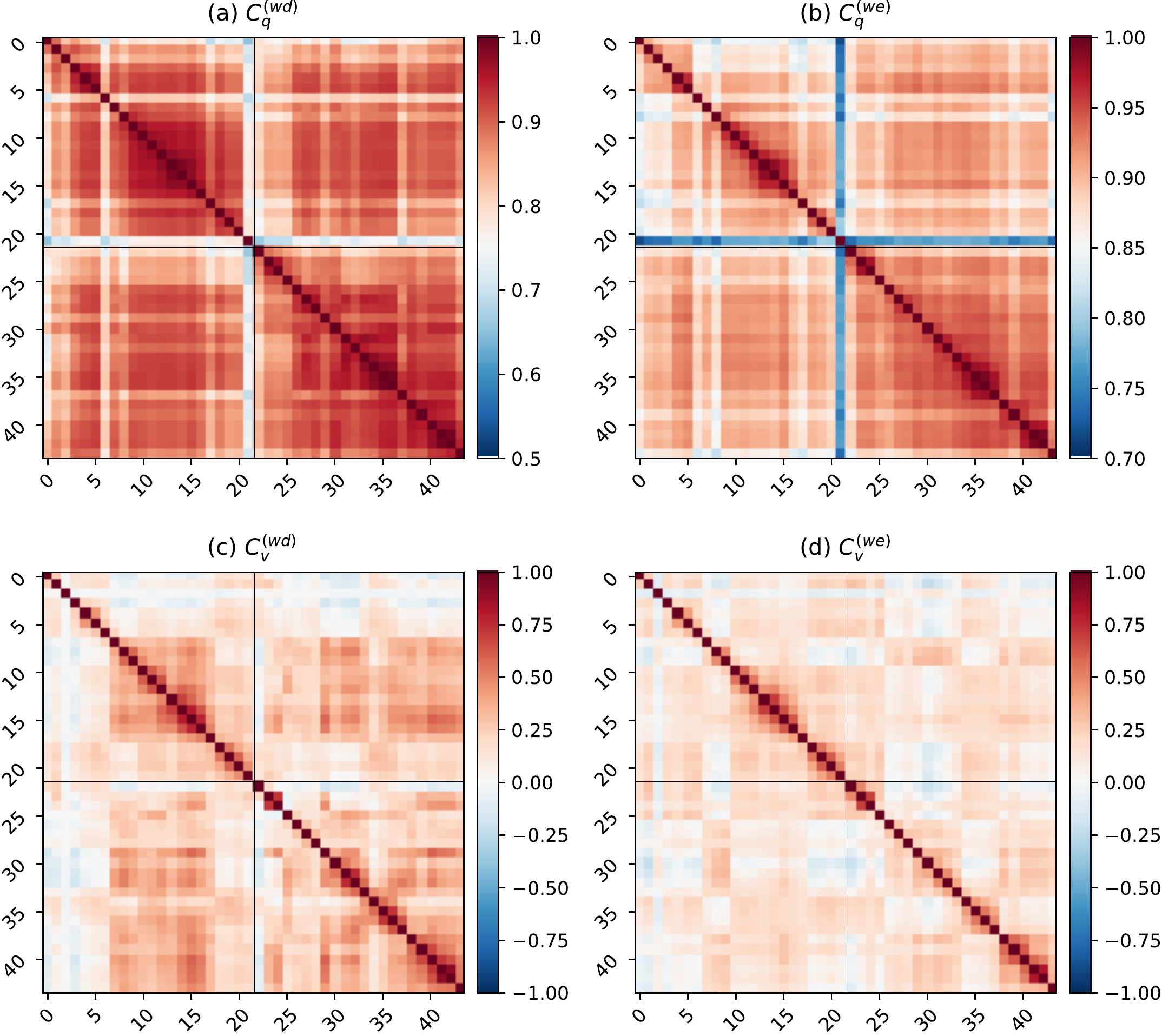}
		\caption{Correlation matrices $C$ of traffic flow (a)-(b) and velocity (c)-(d) for cross-sections on the motorways A40 and A42 averaged over all workdays (wd) and weekends (we) based on data with an aggregation length of 3 minutes. The horizontal and vertical axis indicate the section numbers corresponding to the numbering in Fig. \ref{fig_1_map}. Thin black lines separate the cross-sections of both motorways. For (a) and (b) a different color scale is used to reveal the details of the correlation structure.}
		\label{fig_12_corr_wd_we_3_min}
	\end{center}
\end{figure}

The aggregation time of 15 minutes we applied to our data was chosen based on previous results in reference \cite{Wang2020}. To be able to conduct the investigation of averaged correlation matrices for individual time periods in Section~\ref{sec42} it is necessary to provide more data points. Therefore we aggregated our data to a 3 minute time interval. Since the investigation in Section~\ref{sec42} only introduces correlation matrices for individual time periods, we present the corresponding averaged correlation matrices regarding the whole day in Fig.~\ref{fig_12_corr_wd_we_3_min} for the sake of completeness. A comparison of the correlation matrices resulting from the 15 minute aggregation time in Fig.~\ref{fig_5_corr_wd_we} with the matrices in Fig.~\ref{fig_12_corr_wd_we_3_min} clearly shows that the correlation patterns already manifest in the data set with a 3 minute aggregation time. This is the case for both, traffic flow and velocity correlation matrices. As an example one may consider the diagonal block structures analyzed in the discussion in Section~\ref{sec41}, namely cross-sections 9-16 between Essen and Gelsenkirchen.

\clearpage
\subsection{Distribution of Correlation Coefficients}
\label{ap_distribution}

\begin{figure}[htbp]
	\begin{center}
		\includegraphics[width=0.7\textwidth]{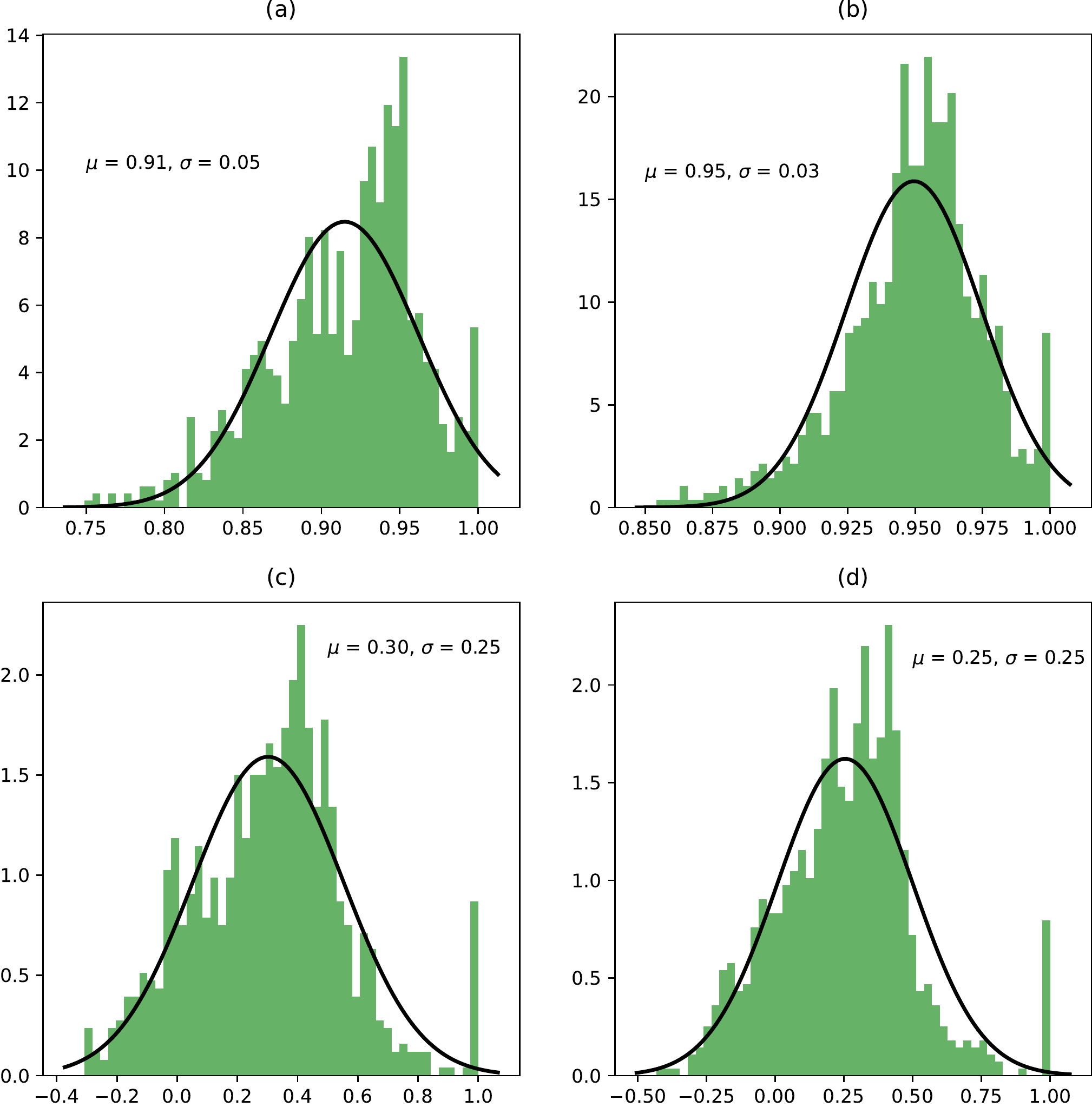}
		\caption{Distribution of correlation coefficients of the Pearson correlation matrices for traffic flows ((a)~and~(b)) and velocities ((c)~and~(d)) on workdays ((a)~and~(c)) and weekends ((b)~and~(d)) from Fig.~\ref{fig_5_corr_wd_we}. Curve fitting was applied to fit Gaussian distribution (black lines).}
		\label{fig_13_corr_wd_we_3_min}
	\end{center}
\end{figure}

For the sake of completeness Fig.~\ref{fig_13_corr_wd_we_3_min} shows the distribution of correlation coefficients of the Pearson correlation matrices in Fig.~\ref{fig_5_corr_wd_we}. We applied curve fitting to fit Gaussian distribution indicated by black lines. Despite the deviations of all four correlation distributions in Fig.~\ref{fig_13_corr_wd_we_3_min} we find that these distributions are close enough to a Gaussian distribution to serve the purpose of our work.

\end{appendices}

\end{document}